\documentclass[%
 reprint,
superscriptaddress,
 amsmath,amssymb,
 aps,
]{revtex4-2}

\usepackage{hyperref}
\hypersetup{
    colorlinks=true,
    linkcolor=blue,
    filecolor=magenta,      
    urlcolor=cyan,
    pdftitle={InAs 2DEG on GaAs substrate},
    }
  
\usepackage[all]{hypcap} 
\usepackage{graphicx}
\usepackage[section]{placeins}
\usepackage{dcolumn}
\usepackage{bm}

\begin{document}

\preprint{APS/123-QED}

\title{Near-surface InAs 2DEG on a GaAs substrate:\texorpdfstring{\\}{ } characterization and superconducting proximity effect}

\author{M\'at\'e S\"ut\H o}
\affiliation{Department of Physics, Institute of Physics, Budapest University of Technology and Economics, M\H uegyetem rkp.\ 3., H-1111 Budapest, Hungary}
\affiliation{MTA-BME Superconducting Nanoelectronics Momentum Research Group, M\H uegyetem rkp.\ 3., H-1111 Budapest, Hungary}

\author{Tam\'as Prok}
\affiliation{Department of Physics, Institute of Physics, Budapest University of Technology and Economics, M\H uegyetem rkp.\ 3., H-1111 Budapest, Hungary}
\affiliation{MTA-BME Superconducting Nanoelectronics Momentum Research Group, M\H uegyetem rkp.\ 3., H-1111 Budapest, Hungary}

\author{P\'eter Makk}
\affiliation{Department of Physics, Institute of Physics, Budapest University of Technology and Economics, M\H uegyetem rkp.\ 3., H-1111 Budapest, Hungary}
\affiliation{MTA-BME Correlated van der Waals Structures Momentum Research Group, M\H uegyetem rkp.\ 3., H-1111 Budapest, Hungary}

\author{Magdhi Kirti}
\affiliation{IOM CNR, Laboratorio TASC, Area Science Park Basovizza, Trieste, 34149, Italy}
\affiliation{Department of Physics, University of Trieste, Trieste, 34128, Italy}

\author{Giorgio Biasiol}
\affiliation{IOM CNR, Laboratorio TASC, Area Science Park Basovizza, Trieste, 34149, Italy}

\author{Szabolcs Csonka}
\email{csonka.szabolcs@ttk.bme.hu}
\affiliation{Department of Physics, Institute of Physics, Budapest University of Technology and Economics, M\H uegyetem rkp.\ 3., H-1111 Budapest, Hungary}
\affiliation{MTA-BME Superconducting Nanoelectronics Momentum Research Group, M\H uegyetem rkp.\ 3., H-1111 Budapest, Hungary}

\author{Endre T\'ov\'ari}
\affiliation{Department of Physics, Institute of Physics, Budapest University of Technology and Economics, M\H uegyetem rkp.\ 3., H-1111 Budapest, Hungary}
\affiliation{MTA-BME Correlated van der Waals Structures Momentum Research Group, M\H uegyetem rkp.\ 3., H-1111 Budapest, Hungary}


\date{\today}

\begin{abstract}
We have studied a near-surface two-dimensional electron gas based on an InAs quantum well on a GaAs substrate. In devices without a dielectric layer we estimated large electron mobilities on the order of $10^5~\mathrm{cm^2/Vs}$. We have observed quantized conductance in a quantum point contact, and determined the g-factor. Using samples with an epitaxial Al layer, we defined multiple Josephson junctions and found the critical current to be gate tunable. Based on multiple Andreev reflections the semiconductor-superconductor interface is transparent, with an induced gap of $\mathrm{125~ \mu eV}$. Our results suggest that this InAs system is a viable platform for use in hybrid topological superconductor devices.
\end{abstract}

\maketitle


\section{\label{sec:intro}Introduction
}

Near-surface two-dimensional electron gases (2DEGs) combined with an epitaxially grown superconducting layer have relatively high mobility, strong spin-orbit coupling and a high-quality superconductor-semiconductor interface, leading to a proximity effect with a hard superconducting gap\cite{Kjaergaard2016a,Suominen2017,Moehle2021}.They represent great promise in the search for a suitable platform for quantum computing\cite{Kitaev2001,Oreg2010,Lutchyn2010,Casparis2018}, and have gathered strong interest over the last few years. The most commonly used material for these quantum wells (QW) is InAs which, in nanowire form\cite{Albrecht2016}, is already established as a potential material for hosting Majorana states. However, a 2D system allows more flexibility in device geometry\cite{Fornieri2019,Dartiailh2021}. 

InAs 2DEGs grown on a GaSb or InP wafer are already well studied and characterized\cite{Tschirky2017, Yuan2020,Wickramasinghe2018}. Compared to GaSb, InP and GaAs are more preferable from a technological point of view, however, an InP substrate is more expensive than GaAs, and less resistive. This is a drawback for quantum computing applications where high-frequency readout and operation of qubits is a major consideration. Superconducting resonators on a highly resistive substrate with low power loss are advantageous, making GaAs a prime candidate. The thickness of InAs QW's based on InP and GaAs is generally limited by strain\cite{Capotondi2005a,Ercolani2008,Hatke2017}, which in turn limits the quality via interface and alloy scattering. 


In this paper we discuss InAs near-surface 2DEGs grown on a GaAs substrate with epitaxial Al on top. The strain is relieved by InAlAs buffer layers\cite{Capotondi2004,Benali2022} between the substrate and the InAs. This allows a QW thickness of 7~nm and, with a 10 nm top InGaAs layer, results in mobilities on the order of $10^5~\mathrm{cm^2/Vs}$, larger than equivalent (not gated) near-surface 2DEGs grown on InP\cite{Wickramasinghe2018, Lee2019, Mayer2019, Yuan2021}. We present the structure of the 2DEG system and the fabrication of devices (Sec.~\ref{sec:fab}), as well as low-temperature transport measurements on Hall bars (Sec.~\ref{sec:transport}) for basic characterization. We demonstrate the operational building blocks required for quasi one-dimensional hybrid nanostructures, such as a quantum point contact (QPC) exhibiting conductance quantization (Sec.~\ref{sec:QPC}) and a Josephson junction (JJ) with tunable supercurrent (Sec.~\ref{sec:JJ}).

\section{\label{sec:fab}Growth, fabrication, and experimental setup}

\begin{figure}
\includegraphics[width=1\columnwidth]{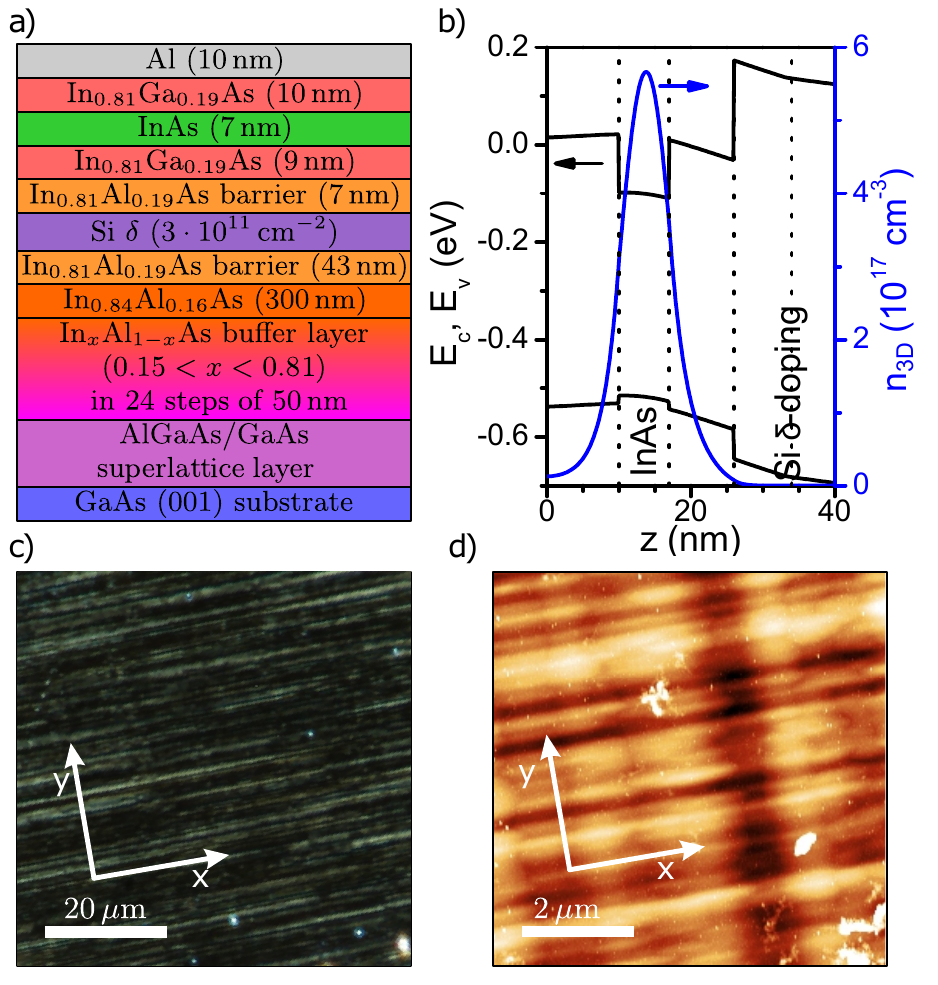}
\caption{\label{fig:layers} a) Schematic of the layer structure. b) Poisson-Schr{\"o}dinger simulation of the conduction and valence band edges relative to the Fermi level ($E_c,~E_v$, black) as a function of depth measured from the surface, and the simulated 3D electron density $n_\mathrm{3D}$ (blue). Dotted lines represent the layer boundaries from panel a, without Al. For the calculation, a background density\cite{Capotondi2004} of $4\cdot 10^{16}~\mathrm{cm^{-3}}$ was assumed in the $\mathrm{In_{0.81}Ga_{0.19}As}$ and $\mathrm{In_{0.81}Al_{0.19}As}$ layers. c) Optical dark field and d) AFM image of the wafer surface. The $x$ and $y$ axes are parallel with the $[\bar110]$ and $[110]$ crystallographic directions, respectively. The RMS surface roughness based on the AFM image is $3.1\,\mathrm{nm}$.}
\end{figure}


The semiconductor heterostructure was realized using solid source molecular beam epitaxy. It was grown on a GaAs (001) substrate, with a strain-relieving step graded buffer of $\mathrm{In_xAl_{1-x}As}$ terminated by a 300~nm thick $\mathrm{In_{0.84}Al_{0.16}As}$ layer for optimal mobility\cite{Benali2022}, as illustrated in Figure~\ref{fig:layers}a. The quantum well was a $\mathrm{7 ~ nm}$ thick InAs layer separated from the wafer surface by a 10~nm thick $\mathrm{In_{0.81}Ga_{0.19}As}$ barrier. Fig.~\ref{fig:layers}b shows a 1D Poisson-Schr{\"o}dinger simulation\cite{Tan1990} of the conduction and valence band edges and the 3D electron density $n_\mathrm{3D}$: the electrons are mainly confined into the local conduction band minimum of InAs. The small thickness of the top InGaAs layer enables finite $n_\mathrm{3D}$ on the surface and therefore forming Ohmic or superconducting contacts to the InAs. To achieve strong superconducting coupling, the heterostructure was covered by $10~\mathrm{nm}$ of Al evaporated in-situ. Following semiconductor growth, the MBE system was idled for about 24 hours to pump residual arsenic, after which the wafer was cooled down to about \text{-}$50^{\circ}\mathrm{C}$, followed by Al deposition at a rate of $0.5~\mathrm{\AA/s}$. Figs.~\ref{fig:layers}c,d show optical dark field and atomic force microscopy (AFM) images of the surface, respectively. A cross-hatched corrugation pattern is visible in both, with features parallel with the $[\bar110]$ and $[110]$ crystallographic directions (arrows labelled $x$ and $y$, respectively). It is the result of the relaxation of dislocations in the InAlAs buffer layer, and is characteristic of metamorphic growth\cite{Capotondi2005}. It can be observed irrespective of the presence of an Al layer.

For transport measurements, Hall bars with a length of 20 or 30~$\mathrm{\mu m}$'s and Josephson junctions (JJs) of a few hundred nm's were fabricated using electron beam lithography (EBL), wet etching, electron beam evaporation and thermal atomic layer deposition (ALD). Here we describe in short the steps for the realization of Hall bars on Al-covered wafers. First, we coated the surface by a layer of adhesion promoter and MMA, and exposed the areas surrounding the Hall bar and electrical leads. The chip was submerged in alkaline MF321 to remove the unmasked Al. The remaining Al served as a hard mask in the second step: the regions not covered by Al were etched\cite{Suominen2017} to a minimum depth of $500~\mathrm{nm}$ by a solution of $\mathrm{H_2O : C_6 H_8 O_7 : H_3 PO_4 : H_2 O_2}$ with a weight ratio of $220:55:3:3$. Etching with MMA or PMMA as a mask was avoided in general as it is accompanied by $\mathrm{\mu m}$-scale underetching. This dopes the 2DEG, as discussed in Section~\ref{sec:QPC}. Next, the Al on the Hall bar (or the JJ weak link) was removed as in the first step, but was left intact elsewhere to serve as metallic or superconducting electrodes. For top gated devices, a 50~nm thick $\mathrm{Al_2 O_3}$ dielectric was grown via thermal ALD at $225\mathrm{^\circ C}$, with precursors of TMA and $\mathrm{H_2 O}$. We note that if the regions exposed in the second step were etched shallower than 500~nm, when later covered by $\mathrm{Al_2 O_3}$, they enabled a lateral leakage current between nominally separated 2DEG mesas, such as actual devices or future top gate bonding pads, similarly to Ref.~\onlinecite{Kjaergaard2015}. The electrical leads of top gates or of devices on wafers lacking an Al layer were fabricated by EBL and Ti/Au evaporation.

In our experiments we have studied devices prepared from wafers with and without an initial Al layer. They were conducted in He-4 systems such as a VTI, and a dilution refrigerator with a base temperature of 20~mK. We used a low-frequency lock-in technique for all measurements except on Josephson junctions, where we used a DAQ (NI USB-6341) to control current and measure voltage for $V(I)$ curves.








\section{\label{sec:results}Results and discussion}

\subsection{\label{sec:transport}Magnetotransport}

Here we discuss magnetotransport on Hall bars, including the estimation of the mobility and scattering times. The results on Hall bars without and with top gates are laid out in separate subsections.

\subsubsection{\label{subsec:transport:bare}Non-gated Hall bars}

First we performed  measurements at 1.5 or 4~K temperature on Hall bars made of wafers without and with an original epitaxial Al layer, in order to determine how the epitaxy and later removal of the Al affects the transport characteristics. The layer structure in these wafers was the same, while the Si $\delta$-doping was varied. 

\begin{figure}
\includegraphics[width=1\columnwidth]{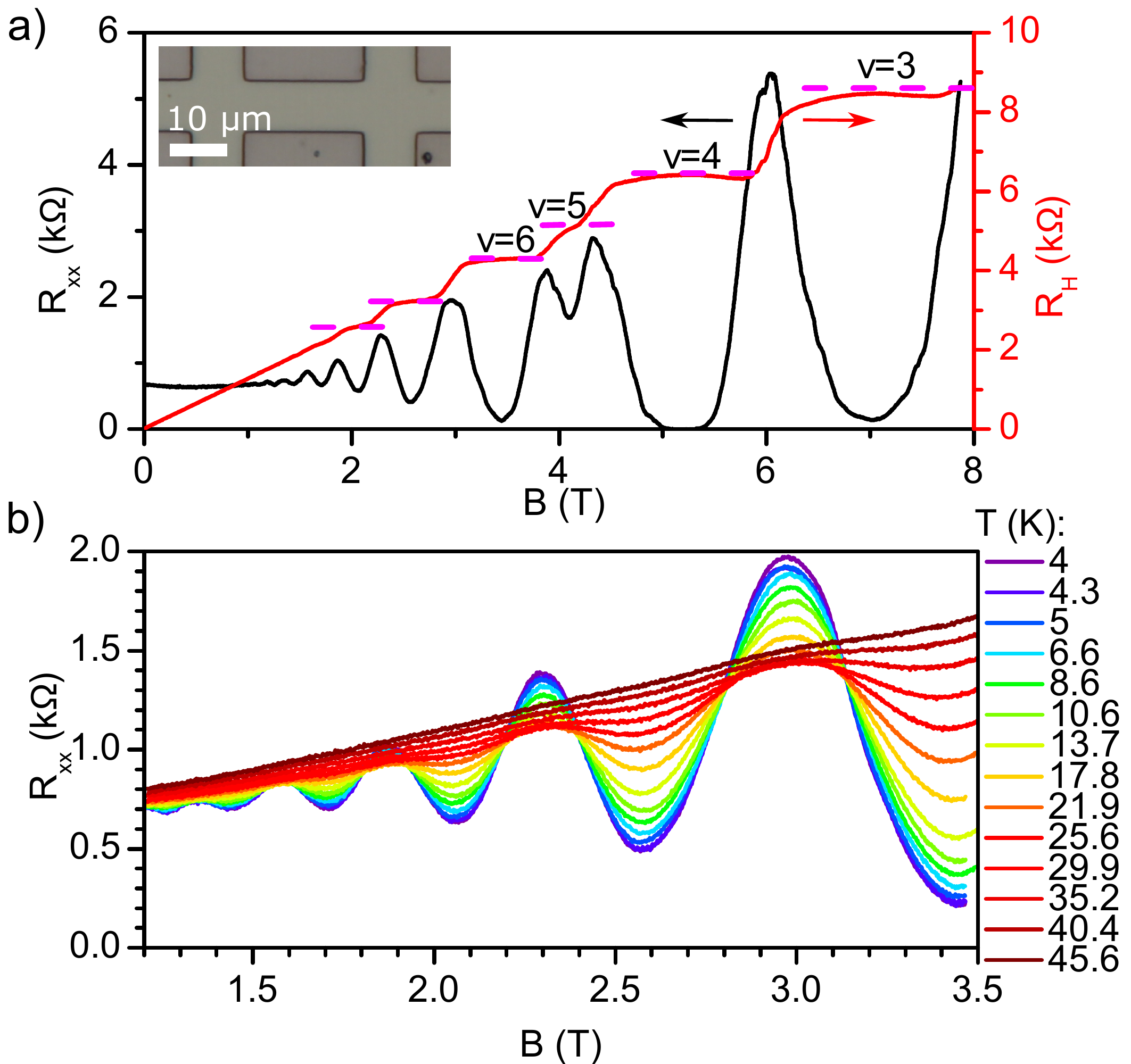}
\caption{\label{fig:SdHO} Magnetotransport on a non-gated Hall bar (sample \#04). a) Longitudinal ($R_{xx}$, black) and Hall ($R_H$, red) resistance as a function of the out of plane magnetic field B at T=1.5~K. The Hall density is $n=4.95\cdot10^{11}$~cm$^{-2}$, the mobility $\mu=86000$~cm$^2/$Vs. Purple dashed lines highlight the expected plateau positions at integer Landau level filling factors $\nu$. Inset: optical photo of a similar Hall bar. b) Longitudinal resistance of the same device at a series of temperatures.}
\end{figure}

In both sets of devices we measured the longitudinal ($R_{xx}$) and Hall ($R_H$) resistance as a function of an out of plane magnetic field $B$. An example is shown in Figure~\ref{fig:SdHO}a of a Hall bar (device \#04) that had originally been covered by Al but which was removed during fabrication. We determined the charge carrier density $n$ and mobility $\mu = \sigma / n e$ from low-field Hall data. Here $e$ is the elementary charge, and $\sigma$ is the conductivity calculated from the resistance $R_{xx}$, Hall bar length $L$ and width $W$ via $ \sigma = 1/\rho_{xx} = L / W R_{xx}$, where $\rho_{xx}$ is the resistivity. In all samples, $n$ and $\mu$ values were consistently around $\left( 4 \mbox{-} 6 \right) \cdot 10^{11} ~\mathrm{cm^{-2}}$ and $ \left( 0.9 \mbox{-} 1.3 \right) \cdot 10^5 ~ \mathrm{cm^2 /Vs}$, respectively. No significant difference was observed in mobility between Hall bars oriented parallel with the $x$ or $y$ directions in Figs.~\ref{fig:layers}c,d. The mean free path $l = \hbar \sqrt{2 \pi n} \mu / e$ was on the order 1~$\mathrm{\mu m}$; here $\hbar$ is the reduced Planck's constant. Moreover, as demonstrated in Fig.~\ref{fig:SdHO}a, Shubnikov-de Haas oscillations (SdHO) could be observed from approximately $B=$1~T which develop into a quantum Hall effect with spin-split 1st and sometimes 2nd Landau levels. These results are irrespective of whether the wafer had an epitaxial Al layer, therefore we conclude that \textit{(i)} this near-surface 2DEG has a remarkably high mobility, and \textit{(ii)} the removal process of Al does not perceivably affect its quality, unlike in Ref.~\onlinecite{Shabani2016}.

In order to gain further information on the properties of these 2DEG devices, we measured the temperature-dependence of the SdHO, as illustrated in Figure~\ref{fig:SdHO}b for device \#04. We collectively fit all the curves in a $B$ range below the onset of the quantum Hall effect using the Lifshitz-Kosevich formula up to the first harmonic\cite{Shoenberg2009},
\begin{equation}
    \delta R = A \frac{x e^{-x_D} }{\sinh \left( x \right)}  \cos\left(2\pi\left(\frac{B_F}{|B|}+0.5\right)\right),
\end{equation}
with $x = 2 \pi^2 k_B T / \hbar \omega_c $ and $ x_D = 2 \pi^2 k_B T_D / \hbar \omega_c $. Here $A$ is the amplitude, $k_B$ is Boltzmann's constant, $T$ is the absolute temperature, and $\omega_c = e |B| / m$ is the cyclotron frequency with the effective mass $m$. $T_D$ is the so-called Dingle temperature, proportional to the disorder broadening of the Landau levels, and $B_F$ is the frequency of the oscillations on the $1/B$ axis, related to the area of the Fermi surface and the density via $B_F = n h / 2 e$ due to spin-degeneracy. In general, the effective mass $m$ determines the oscillations' decay with increasing temperature via the ratio $x$ of the thermal and cyclotron energy. $m$ and $T_D$ determine the oscillations' decay with increasing $x_D \propto 1/B$ due to the overlap of Landau levels with ever smaller spacing. When fitting the curves of $R_{xx}(B)$ at the various temperatures, $A$, $m$ and $T_D$ were shared parameters, but $B_F$ was allowed to vary. For each temperature a 3rd order polynomial was also included to account for the background magnetoresistance. 

For the measurements shown in Figure~\ref{fig:SdHO}b, the fit produced an average $B_F = \left( 10.24 \pm 0.10 \right) \mathrm{T}$, which is consistent with the density estimated from the classical Hall effect, $n=4.95 \cdot 10^{11} ~ \mathrm{cm^{-2}}$. The effective mass in units of the free electron mass $m_e$ was $m/m_e = 0.0281 \pm 0.0001$, which is larger than the bulk value\cite{Mikhailova1996} of 0.023. This may be due to strain or the wavefunction extending into the InGaAs layers which have a larger $m$ (Ref.~\onlinecite{Lin-Chung1993} and Fig.~\ref{fig:layers}b). The Dingle temperature was $T_D = (17.5 \pm 0.1)~\mathrm{K}$, from which we can calculate the elastic scattering time, i.e. the lifetime of the Bloch waves\cite{Monteverde2010}: $\tau_e = \hbar / 2 \pi k_B T_D \approx 0.069~ \mathrm{ps}$. In contrast, the transport lifetime related to backscattering which appears in the mobility is $\tau_\mathrm{tr} = \mu m / e \approx 1.37 ~ \mathrm{ps}$, approximately 20 times larger. This indicates that backscattering is rare, and long-range scatterers dominate: a further sign of the high quality of the 2DEG. In similar measurements over seven other samples, we consistently found values of $m/m_e$ in the range 0.025-0.03, while $T_D$ was between 17-28~K, with one outlier of 53~K. These values of $m$ and $T_D$ are lower than in a similar quantum well structure\cite{Yuan2020}, which is likely due to the thicker InAs well (7~nm here instead of 4~nm), the additional barriers between the step graded buffer and the well, and the lack of a gate dielectric. 

\subsubsection{\label{subsec:transport:gated}Hall bars with a global top gate}

\begin{figure}
\includegraphics[width=1\columnwidth]{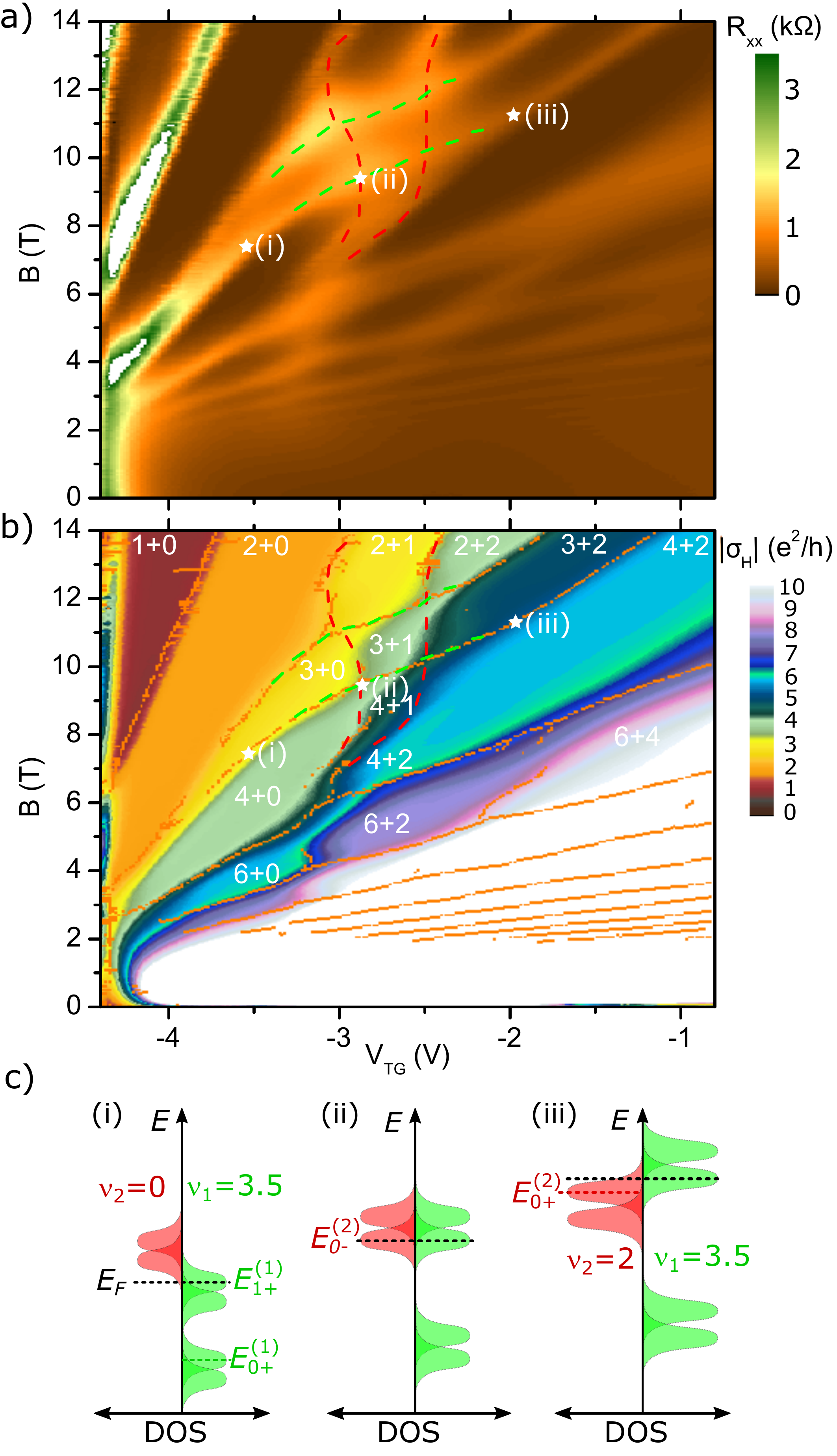}
\caption{\label{fig:top_gate_VBmap} Magnetotransport on top gated Hall bar \#11. a) Resistance map as a function of $V_\mathrm{TG}$ and $B$ at 4~K. The dashed lines highlight crossings of the Landau levels of the 1st (green) and 2nd (red) subbands. b) The corresponding map of the Hall conductivity. Orange dots indicate the resistance maxima in a), while white numbers are filling factors of the subbands on quantum Hall plateaus: $\nu_1 + \nu_2$. c) Landau level density of states (DOS) illustrations for the three stars on the lower green dashed line in panels a,b, i.e. keeping $\nu_1 = 3.5$ constant and $E_F = E_{1 +}^{(1)}$ (highlighted by a black dashed line in all subpanels).
}
\end{figure}

Next we studied the transport properties of Hall bars equipped with a $50~\mathrm{nm}$ thick $\mathrm{Al_2 O_3}$ dielectric layer and a global top gate electrode, as a function of a perpendicular magnetic field and top gate voltage $V_\mathrm{TG}$. The first device we discuss is \#11, of which Figs.~\ref{fig:top_gate_VBmap}a,b present colormaps of the longitudinal resistance $R_{xx}$ and the Hall conductivity $\left| \sigma_H \right| = \left| R_H \right| / \left( R_H^2 + \rho_{xx}^2 \right) $, respectively. The voltage threshold where the 2DEG is depleted is approximately around $V_\mathrm{th} = -5.1~\mathrm{V}$ (not shown). A Landau fan can be observed up to $-3.5~ \mathrm{V}$ as lines of $R_{xx}$ maxima originating in $V_\mathrm{th}$ and $B=0$~T. On these resistance ridges the Landau levels (LLs) are half filled, i.e. the filling factor $\nu$ is odd or (if spin-split) half-integer. As we follow a ridge, when increasing $B$ by a small amount, $\nu$ must remain constant: we keep pace with the increasing degeneracy of the LLs by increasing $n$ via $V_\mathrm{TG}$ with a rate of $\mathrm{d}B / \mathrm{d}V_\mathrm{TG} = C_\mathrm{TG} h / e \nu$ where $C_\mathrm{TG}$ is the capacitance in units of [m$^{-2}$/V]. In Fig.~\ref{fig:top_gate_VBmap}b, the resistance maxima are represented by orange dots.

However, around -3.5~V and $n\approx 6\cdot 10^{11} ~\mathrm{cm^{-2}}$, the lines of the fan seem to break at low fields and continue with different slopes $\mathrm{d}B / \mathrm{d}V_\mathrm{TG}$. Moreover, a second Landau fan appears, which corresponds to LLs of a 2nd subband\cite{Ellenberger2006,Pauka2020}. Fig.~\ref{fig:layers}b shows the calculated conduction band edge at a slightly lower density of $n=5\cdot 10^{11} ~\mathrm{cm^{-2}}$: here the Fermi-energy ($E=0$) is already close to the top of the InAs quantum well. Therefore we attribute the 2nd subband to the ground state in the $26~\mathrm{nm}$ wide secondary quantum well formed by the InGaAs and InAs layers and bordered by InAlAs. 

In the quantum Hall regime the physics of a two-subband system can be understood from the union of the two sets of LLs\cite{Ensslin1993}. If the Fermi level $E_F$ is far from any LL, i.e. the individual filling factors of the subbands $\nu_{1,2}$ are both near integers, $R_{xx}$ is zero, while in $ \sigma_H $ there is a quantum Hall plateau at $\nu_\mathrm{tot} e^2 / h$ using the total filling factor $\nu_\mathrm{tot} = \nu_1 + \nu_2$. The white numbers in Fig.~\ref{fig:top_gate_VBmap}b denote $\nu_1 + \nu_2$, which correspond well to the experimentally observed plateau values. Nevertheless, the resistance and Hall conductivity in the two-subband regime ($V_\mathrm{TG} > -3.5~ \mathrm{V}$ in Fig.~\ref{fig:top_gate_VBmap}) are not simple superpositions of two Landau fans and their sets of quantum Hall plateaus, and the determination of $\nu_{1,2}$ is not straightforward. Near expected LL crossings at high $B$ the resistance ridges follow curved lines. The most visible examples are highlighted by pairs of dashed lines: they follow the 1st subband's spin-split first LL (green) and the 2nd subband's spin-split zeroeth LL (red). We denote a spin-split Landau level of subband (j) with quantum number $n \geq 0$ by $E_{n \pm}^{(j)} = \hbar \omega_c^{(j)} (n+1/2) \pm E_\mathrm{Zeeman} + E^{(j)}$ where the $+$ and $-$ represent spin orientation relative to $B$, $E^{(j)}$ is a subband-dependent offset, and $\omega_c^{(j)}$ is the subband-dependent effective cyclotron frequency: $\omega_c^{(1)} > \omega_c^{(2)}$ is expected due to the larger $m$ in InGaAs. Then these four lines correspond to the half-filled levels $E_{1 \pm}^{(1)}$ (green: $\nu_1$=2.5,~3.5) and $E_{0 \pm}^{(2)}$ (red: $\nu_2 $=0.5,~1.5).

To understand why the lines curve, let us follow the points highlighted by white stars along the lower green dashed line: $E_F = E_{1 +}^{(1)}$ and $\nu_1 = 3.5$ if we neglect the overlap with its spin-split pair $E_{1 -}^{(1)}$. As illustrated by the LL peaks in the density of states in Fig.~\ref{fig:top_gate_VBmap}c, point (i) is far from a crossing, therefore $\nu_2 = 0$ and $E_F < E_{0 -}^{(2)}$, and this resistance ridge has a slope of $\mathrm{d}B / \mathrm{d}V_\mathrm{TG} = C_\mathrm{TG} h / e \nu_\mathrm{tot}$. As we increase $B$, $ E_{1 +}^{(1)}$ increases with a higher rate than $E_{0 -}^{(2)}$ due to the larger quantum number and cyclotron frequency, and the spin orientation. Therefore, nearing point (ii), $E_F = E_{1 +}^{(1)}$ approaches the disorder-broadened $E_{0 -}^{(2)}$. Following the line while increasing $B$ now requires more density input as we must simultaneously fill both LLs. This corresponds to a smaller slope $\mathrm{d}B / \mathrm{d}V_\mathrm{TG}$, as can be observed in panels a,b: $\nu_\mathrm{tot}$ increases from approximately 3.5 (i) to 4 (ii). After crossing both red dashed lines, $E_F = E_{1 +}^{(1)} > E_{0 \pm}^{(2)}$ and the resistance ridge becomes linear again but with a smaller slope (point (iii)), since $\nu_\mathrm{tot}$ has increased to 5.5. In contrast, following either red dashed line while increasing $B$ would mean a decrease in $\nu_\mathrm{tot}$: emptying a LL produces a nearly vertical trend here, possibly with a negative slope. If the disorder-related broadening of a LL was smaller, the crossings on a $V_\mathrm{TG}$--$B$ map would be sharper\cite{Ellenberger2006,Tschirky2017}. The indicated values of $\nu_1 + \nu_2$ in Fig.~\ref{fig:top_gate_VBmap}b were chosen to be consistent with plateaus in $|\sigma_H|$ and the observed crossings. 



\begin{figure}
\includegraphics[width=1\columnwidth]{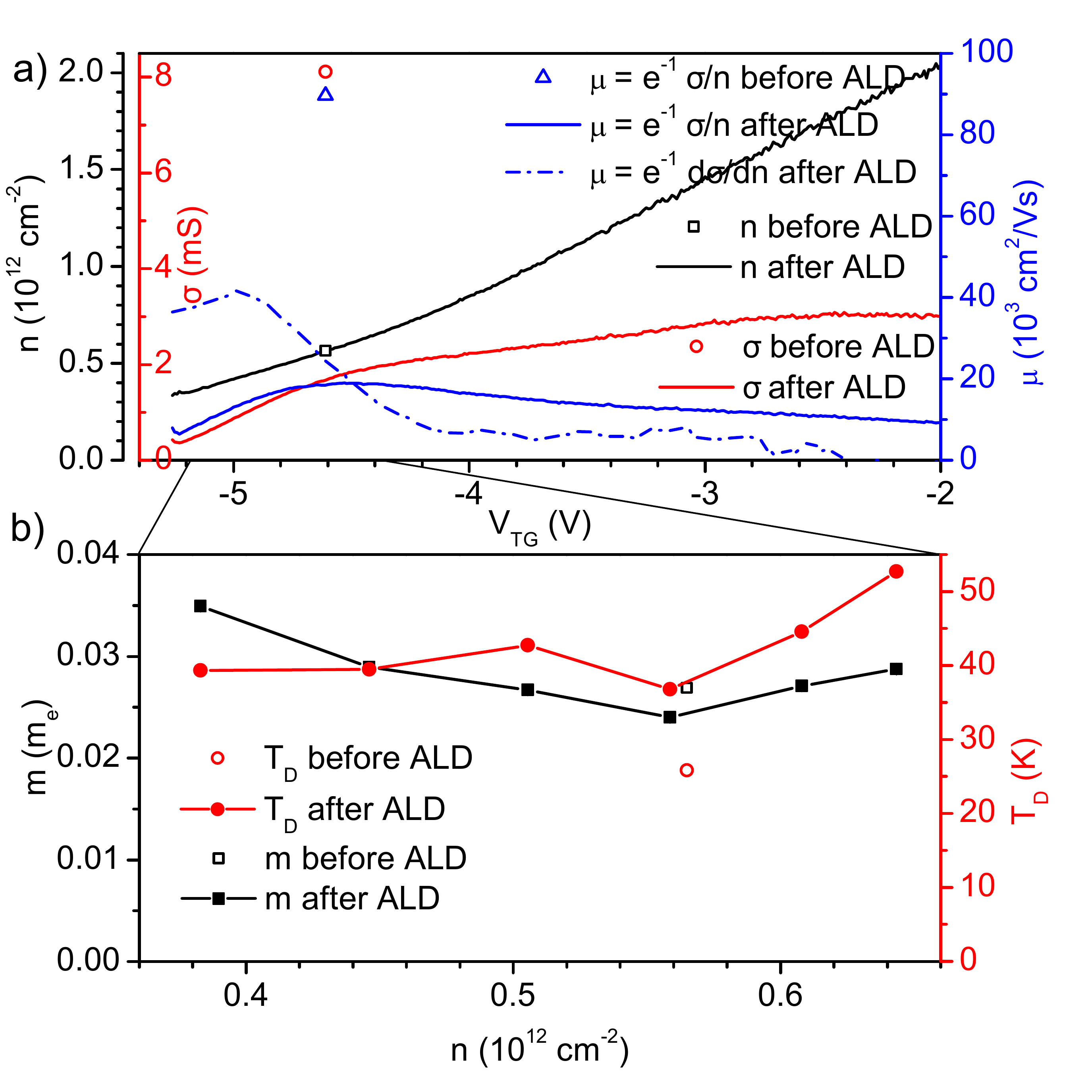}
\caption{\label{fig:top_gate_1} Characterization of top gated Hall bar \#14. a) Density $n$ (black), conductivity $\sigma$ (red) and mobility $\mu$ (blue) at 1.5 K. Single data points are values before ALD, while solid or dashed lines are measurements as a function of top gate voltage $V_\mathrm{TG}$ after ALD and gate electrode fabrication. The abscissa of the single points was chosen as the conjunction of the density data. b) Effective mass $m$ and Dingle temperature $T_D$ as a function of density $n$. Single data points are values before ALD.
}
\end{figure}

In the following, we will discuss top gated device \#14 which we studied more extensively, but in lower magnetic fields compared to \#11. It was first cooled down before ALD and measured at 1.5~K: the results of low-field magnetotransport ($n$, $\sigma$, $\mu$) and Lifshitz-Kosevich analysis of SdHO ($m$, $T_D$) are shown as single datapoints in Fig.~\ref{fig:top_gate_1}a,b, respectively. After ALD but before top gate deposition it was cooled down again for measurement, however, it was found to be insulating. This suggests that the $\mathrm{Al_2 O_3}$ deposition effectively p-doped the sample, possibly by pinning the Fermi level mid-gap. After top gate fabrication, the sample was again conductive at $V_\mathrm{TG}=0 ~ \mathrm{V}$. We measured a $V_\mathrm{TG}$-$B$ map of the resistance up to 8~T (not shown), which is qualitatively similar to sample \#11 in Fig.~\ref{fig:top_gate_VBmap}, including the onset of a 2nd Landau fan around $V_\mathrm{TG}=-4.4 ~ \mathrm{V}$. The depletion threshold is around $V_\mathrm{th}=-5.3$~V.

The curves of $n$, $\sigma$, and $\mu$ based on low-$B$ data as a function of $V_\mathrm{TG}$ are shown in Fig.~\ref{fig:top_gate_1}a. Compared to the pre-ALD values at $n=5.65\cdot10^{11} ~ \mathrm{cm^{-2}}$, the ALD process and top gate fabrication has significantly reduced the conductivity and mobility. If we extrapolate the left ends of the $n \left( V_\mathrm{TG} \right)$ and $\sigma \left( V_\mathrm{TG}\right)$ curves in Fig.~\ref{fig:top_gate_1}a, they intercept the horizontal axis at different points (not shown), similarly to Ref.~\onlinecite{Pauka2020}. Moreover, based on two-terminal conductance curves for all possible contact pairings, we found that the threshold voltages where individual contacts open differ by several tenths of Volts. These suggest that doping is inhomogeneous, and that the mobilities, calculated either as $\mu = e^{-1} \sigma / n$ or $\mu = e^{-1} \mathrm{d}\sigma / \mathrm{d}n$, are effectively average values. 


We also studied the temperature-dependence of SdHO at several gate voltages in the single-subband regime for Lifshitz-Kosevich analysis. The Dingle temperatures shown by the connected red dots in Fig.~\ref{fig:top_gate_1}b are larger than before ALD (single red dot), confirming the increased disorder. The effective mass (black) was only weakly affected by ALD. There is no significant trend in the density dependence of either $m$ or $T_D$. Therefore we can estimate the lowest energy of the second subband relative to the first: at its onset of $-4.4~\mathrm{V}$, $n \approx 6.5 \cdot 10^{11}~ \mathrm{cm^{-2}}$, and using $m/m_e = 0.028$ we get $E^{(2)} - E^{(1)} = \hbar^2 \pi n / m \approx 56 ~ \mathrm{meV}$. This is comparable to the approximately $70~\mathrm{meV}$ between the bottom of the InAs quantum well and the minimum of the conduction band near $z=26~\mathrm{nm}$ in Fig.~\ref{fig:layers}b. For a more accurate comparison, the zero-point energies of the size quantized states in the primary and secondary quantum wells should also be taken into account.

From the onset of the 2nd subband, $V_\mathrm{TG} > -4.4 ~ \mathrm{V}$, the slope of $n \left( V_\mathrm{TG} \right)$ increases (see Fig.~\ref{fig:top_gate_1}a), while the slope of $\sigma \left( V_\mathrm{TG} \right)$ decreases, corresponding to a drop in the mobility $\mu \left( V_\mathrm{TG} \right) = e^{-1} \sigma / n$. We attribute the change in the slope of $n \left( V_\mathrm{TG} \right)$ to the wavefunction of the 2nd subband extending well into the InGaAs barriers on the sides of the InAs QW (Fig.~\ref{fig:layers}b), leading to an increased geometrical capacitance. We note that with multiple-carrier conduction as is the case here, the plotted density $n$ based on the classical Hall effect is not necessarily the total of the subband densities $n_{1,2}$, while $\mu$ calculated from $n$ is not the mobility of electrons in either subband ($\mu_{1,2}$), unless the latter are close\cite{Stormer1982}, $\mu_1 \approx \mu_2$. Based on the constant slope of $R_H \left( B \right) $ in the low-$B$ regime, $\mu_1 \approx \mu_2 \approx \mu$ is a good assumption, therefore $n \approx n_1 + n_2$, and its increased slope can indeed be explained as above. As for the behavior of $\sigma$ and $\mu$, the open 2nd subband makes inter-subband scattering possible\cite{Stormer1982,Ensslin1993,Pauka2020}: this is expected to reduce $\mu$ and decrease the slope of $\sigma \left( V_\mathrm{TG} \right)$, which is in accordance with our observations. Even so, we estimate that the mean free path $l$ is over 200~nm for most of the gate voltage range. We also performed Lifshitz-Kosevich analysis of 1st-subband SdHO in the two-subband regime, which failed due to the interference of a few oscillations from the 2nd subband.


We studied further top gated Hall bars, all of which exhibited similar trends in $\sigma$, $n$ and $\mu$ as a function of $V_\mathrm{TG}$, with a peak $\mu = e^{-1} \sigma / n$ around $\left( 1.5 \mbox{-} 2 \right) \cdot 10^4 ~ \mathrm{cm^2/Vs}$, and a large density at $V_\mathrm{TG} = 0 ~ \mathrm{V}$. The behavior of device \#11 of Fig.~\ref{fig:top_gate_VBmap} was also comparable, though $\sigma$ and $\mu$ (and therefore $l$) were approximately twice as large as those of sample \#14 in Fig.~\ref{fig:top_gate_1}, with $\mu$ reaching $5 \cdot 10^4 ~ \mathrm{cm^2/Vs}$ and $l\approx 0.65~\mathrm{\mu m}$. 

\begin{figure*}[bht!]
\includegraphics[width=1\textwidth]{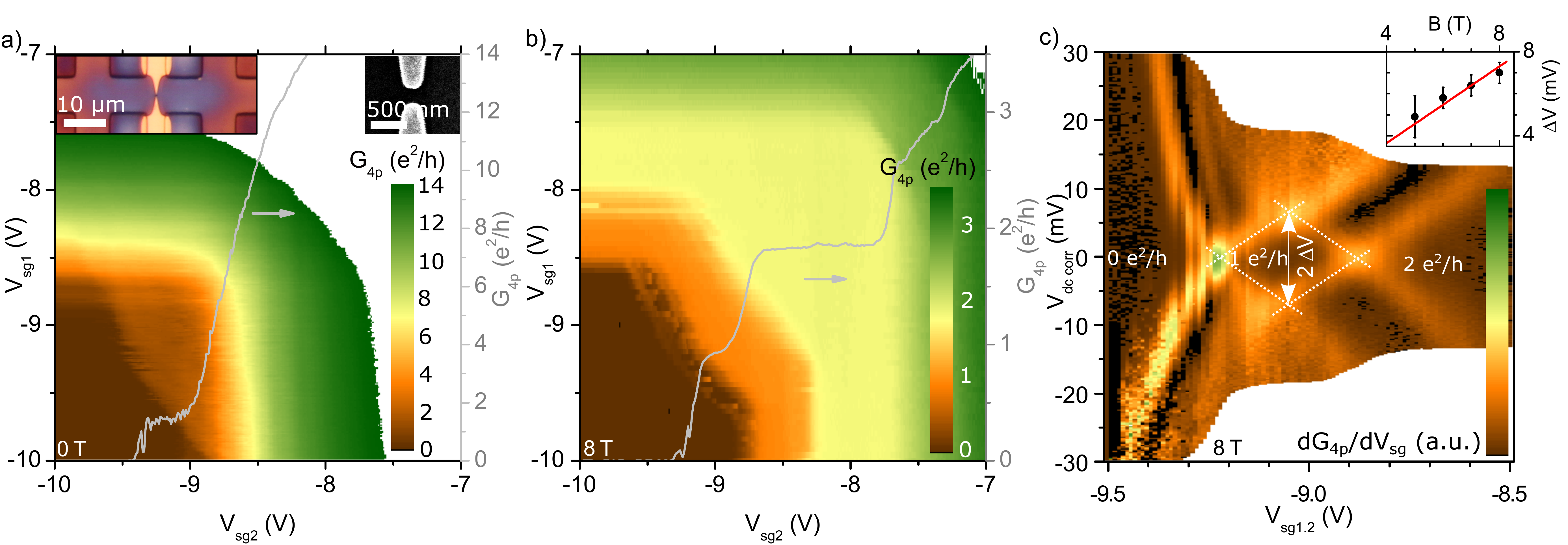}
\caption{\label{fig:qpc}Transport in a quantum point contact. a) Four-point conductance $G_{4p}$ as a function of split gate voltages $V_\mathrm{sg1,2}$ at 1.5~K and 0~T. The grey curve is the conductance along the diagonal $V_\mathrm{sg1}=V_\mathrm{sg2}$ (see the right vertical axis). Insets: optical microscope image of the sample (left), and scanning electron micrograph of a similar pair of split gates (right). b) An equivalent map of $G_{4p} \left( V_\mathrm{sg1},V_\mathrm{sg2} \right)$ at $B=8$~T. The grey curve is a diagonal cut. c) Bias spectroscopy along the diagonal of the gate-gate map at 8~T, showing the numerical derivative $\mathrm{d}G_{4p} / \mathrm{d} V_\mathrm{sg1,2}$. $V_{\mathrm{dc~ corr}}$ is the estimated DC voltage drop on the QPC. Inset: half-size $\Delta V$ of the $1~e^2/h$ plateau as a function of $B$.
}
\end{figure*}

To summarize this section, the removal of the epitaxial Al does not affect the properties of the 2DEG, while fabricating a gate dielectric and a top gate introduces inhomogeneity and a strong n-doping. Nevertheless, we expect that the mean free path value of hundreds of nm's in top gated samples is adequate to realize a QPC.

\subsection{\label{sec:QPC}Quantum point contact measurements}

Here we discuss measurements on a Hall bar device equipped with a pair of top split gates (sg) that were prepared in the middle to define a constriction, as shown by the optical micrograph in the left inset of Figure~\ref{fig:qpc}a. The separation of the split gates is estimated to be 140~nm. For the measurements, an AC bias voltage $V_{2p} = 100 ~ \mathrm{\mu V}$ was applied between the ends of the Hall bar (source and drain), the four-point voltage $V_{4p}$ was measured between side contacts, while the AC current $I$ was measured between the source and the drain contacts. 

Figure~\ref{fig:qpc}a shows the four-point differential conductance $G_{4p}=I/V_{4p}$ at 1.5~K as a function of the gate voltages $V_\mathrm{sg1,2}$. In the bottom left region $G_{4p}=0$: the constriction and the areas under the split gates are all insulating. Increasing either voltage while leaving the other a constant $-10~\mathrm{V}$ opens conduction in the area under the corresponding gate. In the vicinity of the diagonal $V_\mathrm{sg1}=V_\mathrm{sg2}$ (see also the grey curve) a plateau of $G_{4p} \approx 2~e^2/h$ can be observed. On the gate-gate map the outline of this feature has a negative slope, $\mathrm{d}V_\mathrm{sg1} / \mathrm{d}V_\mathrm{sg2} < 0$, indicating that it is tuned by both gates and confirming that it originates in the constriction. As we follow the diagonal, higher order plateaus cannot be observed since the areas under the split gates become conductive. This suggests the potential well is shallow, which we attribute to the narrowness of the constriction. The plateau value is smaller than $2~e^2/h$ due to the serial resistance contribution of the diffusive 2DEG areas on the left and right of the QPC (see insets). We have mentioned in Section~\ref{subsec:transport:gated} that areas covered by $\mathrm{Al_2 O_3}$ but without a top gate were insulating. In contrast, this Hall bar was conductive, which we attribute to using a PMMA mask at a mesa etching step instead of Al. The etching solution crept under the mask, leading to the discoloration in the left inset in Fig.~\ref{fig:qpc}a and an apparent n-doping of the 2DEG.


Figure~\ref{fig:qpc}b shows the same conductance map at 8~T out of plane magnetic field, well inside the quantum Hall regime: here QPC physics is expected to be overshadowed by Landau levels and edge states. On the diagonal cut (grey curve), plateaus of $G_{4p}$ can be observed near 1 and 2~$e^2/h$. Based on Hall measurements, the filling factor of the outside regions at this field is $\nu \approx 4$. Therefore, as long as the filling factor in the QPC (tuned by $V_\mathrm{sg1,2}$) is smaller, not all edge states are transmitted, and this is effectively a two-terminal quantum Hall measurement, hence the finite plateaus in $G_{4p}$.

We have determined the effective g-factor $g_{\perp}$ in an out of plane magnetic field $B$ by performing bias spectroscopy on the first plateau, $G_{4p} = 1 ~ e^2/h$ with $V_\mathrm{sg1}= V_\mathrm{sg2}$, at multiple values of $B$. An example is shown in Fig.~\ref{fig:qpc}c, where we plot $\mathrm{d} G_{4p} / \mathrm{d} V_\mathrm{sg1,2}$ for better visibility. Here the DC and AC voltage biases were applied simultaneously. To get the DC voltage drop on the QPC, $V_\mathrm{dc}$ was corrected for a serial resistance contribution. This resistance originates in the contacts and potentially in the regions outside the QPC as well, since the filling factor outside the QPC is not necessarily an integer at arbitrary $B$. We estimated the serial resistance contribution $R_c$ at each $B$ field by comparing the zero-bias plateau values in $G_{2p}=I/V_{2p} = 1/R_{2p}$ to $1$ and $2~e^2/h$. The corrected voltage drop was calculated at each point of the map using a smoothed zero-bias $R_{2p} \left( V_\mathrm{sg1} = V_\mathrm{sg2} \right)$ curve via $V_\mathrm{dc~ corr} = V_\mathrm{dc} \cdot \left( R_{2p} -R_c \right) / R_{2p} $. For instance, in the regime where the QPC is closed, all of the voltage drops on it and $R_{2p}$ diverges, therefore the voltage drop is the applied voltage. At maximum voltage, this means $V_\mathrm{dc ~ corr} = V_\mathrm{dc} = 30~\mathrm{mV}$. In contrast, when on the 1st plateau, the maximum is $V_\mathrm{dc ~ corr} \approx 20 $~mV due to $R_c \approx 17 ~ \mathrm{k \Omega}$.

The 1st plateau on the gate-bias map in Figure~\ref{fig:qpc}c is shaped like a diamond, as highlighted by the dashed lines. Since the spin-split Landau levels have the same orbitals, its half-size $\Delta V$ is equal to the Zeeman energy in eV's, which is plotted in the inset for a set of $B$ fields. For decreasing $B$ the splitting decreases, but the diamonds become less well-defined, and eventually become indiscernible from a smooth background. Since $e \Delta V =  \left| g_{\perp} \right| \mu_B B$ where $\mu_B$ is the Bohr-magneton, with a linear fit that crosses the origin (red line in the inset) we get $ \left| g_{\perp}  \right| = 15.8 \pm 0.4$. This is closer to the InAs bulk value\cite{Pidgeon1967,Konopka1967} of $\sim 15$ than those reported in similar structures on InP or GaSb substrates in Refs.~\onlinecite{Lee2019, Mittag2019}.

\subsection{\label{sec:JJ}Josephson junction measurements}

 \begin{figure*}[!htbp]
\includegraphics[width=1\textwidth]{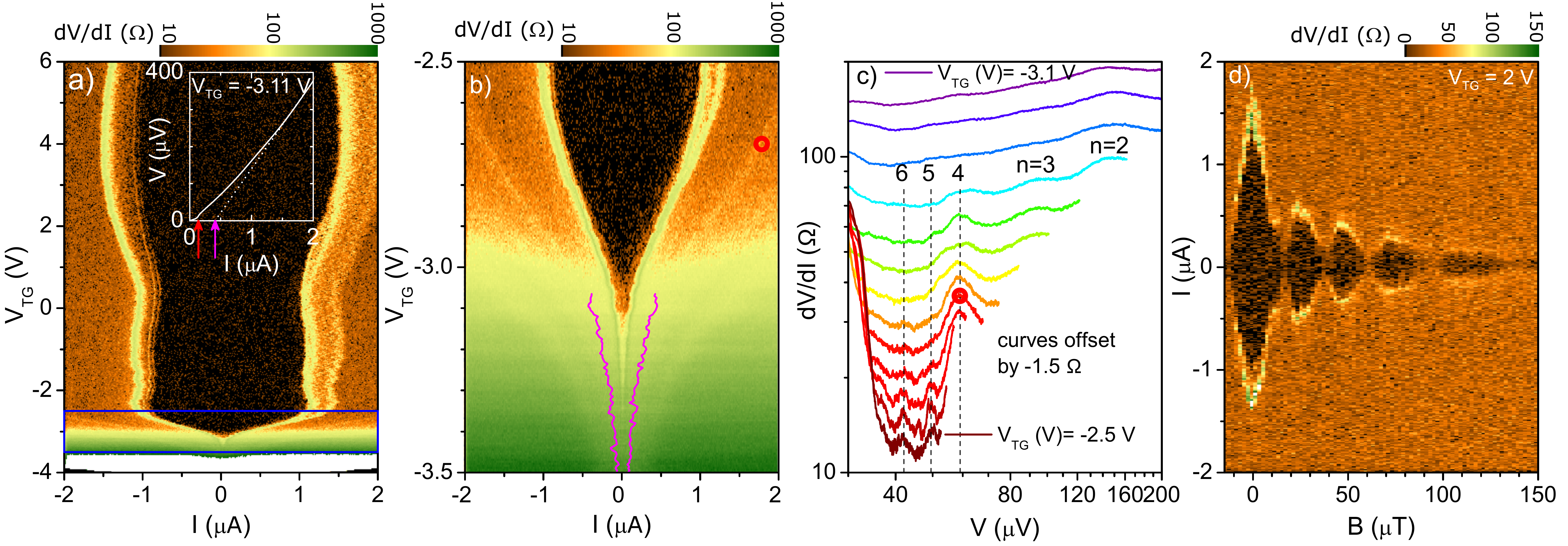}
\caption{\label{fig:JJ} Measurements on a top gated Josephson junction. a) The resistance $\mathrm{d}V / \mathrm{d} I$ on a logarithmic scale as a function of the applied (DC) current $I$ and $V_\mathrm{TG}$. Inset: an example $V(I)$ curve (solid line) at $V_\mathrm{TG}=-3.11~\mathrm{V}$. The dotted line is the fit to the high-$|V|$ part. Red and purple arrows denote the critical current $I_c$ and excess current $I_\mathrm{exc}$, respectively. b) A zoom of the low-density region as highlighted by the blue rectangle in panel a. Purple curves plot the estimated $I_\mathrm{exc}$. c) Smoothed $\mathrm{d}V / \mathrm{d} I$ data plotted versus $V$ at a series of gate voltages increasing from $V_\mathrm{TG}=-3.1~\mathrm{V}$ (purple line) with 0.05~V step size to $V_\mathrm{TG}= -2.5~\mathrm{V}$ (dark red). The red circle is equivalent to the point highlighted in panel b) at $V_\mathrm{TG}=-2.7~\mathrm{V}$. d) Fraunhofer pattern as a function of out of plane magnetic field $B$ at $V_\mathrm{TG}=2~\mathrm{V}$.
}
\end{figure*}

We fabricated multiple Josephson junctions by removing a thin line of Al across a stretch of 2DEG mesa. Here we present data on a $L=0.3~\mathrm{\mu m}$ long and $W=9~\mathrm{\mu m}$ wide Josephson junction (JJ) equipped with a global top gate, collected in a dilution refrigerator with a base temperature of 20~mK. Over a range of gate voltages, we measured a hundred $V(I)$ curves at each $V_\mathrm{TG}$ by sweeping the DC current bias $I$, where $V$ is the voltage drop on the JJ. After averaging at each $V_\mathrm{TG}$, we plot the numerical derivative $\mathrm{d}V / \mathrm{d} I$ in Fig.~\ref{fig:JJ}a. Panel b of the same figure shows a zoom at low gate voltages. We define the critical current $I_c$ as the point of maximal resistance $\mathrm{d}V / \mathrm{d} I$ on either side of the zero-resistance (black) region. The critical current is clearly tunable with the gate voltage: it quickly increases from zero then, after reaching $1.1~\mathrm{\mu A}$ around $V_\mathrm{TG} = -2.6 ~ \mathrm{V}$, it is only weakly affected by $V_\mathrm{TG}$. This point approximately coincides with a change in the slope of the normal state conductance $G_N = dI/dV$ (estimated at $I=-1.8~\mathrm{\mu A}$, not shown). Similarly to our other top gated samples such as device \#14 in Fig.~\ref{fig:top_gate_1}, this is probably the point where the 2nd subband opens, and inter-subband scattering becomes possible. While $I_c$ varies over a wide range and reaches up to $1.5~\mathrm{\mu A}$, the product $I_c \cdot R_N$ (using the normal state resistance $R_N $, calculated at $I=-1.8~\mathrm{\mu A}$) is practically independent of the gate voltage, and varies around 30~$\mathrm{\mu V}$. 

In Fig.~\ref{fig:JJ}b, a zoom of panel a, local maxima can be observed for $|I| > |I_c|$, such as highlighted by the red circle at $V_\mathrm{TG}=-2.7~\mathrm{V}$. When we plot horizontal resistance cuts as a function of $V$ instead of $I$, as is demonstrated in Fig.~\ref{fig:JJ}c, the positions of local maxima coincide between the curves. We attribute these to multiple Andreev reflections (MAR). Normally, for a superconductor-insulator-superconductor junction, one can observe dips in the resistance due to MAR at points defined by $V=2\Delta / n e$ with $n=1,2,3,...$ where $\Delta$ is the superconducting gap induced in InAs by the proximity effect. However, in junctions with a normal metal or semiconductor, high transmission probability can be achieved\cite{Kjaergaard2017}, leading to local maxima in $\mathrm{d}V / \mathrm{d} I$, as is the case here. We denote these points by numbers and dashed lines in Fig.~\ref{fig:JJ}c. Therefore, our results suggest that the superconductor-semiconductor interface here is relatively transparent, and estimate the induced gap: $\Delta = \left( 125 \pm 3 \right)~\mathrm{\mu e V}$. In comparison, the gap of bulk Al is $\Delta_\mathrm{Al} \approx 170~\mathrm{\mu e V}$\cite{Kittel1996}, while the gap of thin Al films can vary between 200--240~$\mathrm{\mu e V}$\cite{Shabani2016, Kjaergaard2017, Mayer2019,Lee2019,Yuan2021}. The discrepancy between $\Delta$ estimated from MAR and $\Delta_\mathrm{Al}$ is due to the fact that the wafer is contacted by the Al layer from the top, resulting in a smaller, induced gap $\Delta$ in the 2DEG underneath, and MAR processes at the edges of the JJ are sensitive to the latter\cite{Kjaergaard2017,Baumgartner2021}.


Another characteristic value of a JJ is the excess current $I_\mathrm{exc}$, which reflects the conductance enhancement of the junction via Andreev processes. We define it as the  $V=0$ axis intercept of linear fits to the section of a $V(I)$ curve where $|V| \gg \Delta/e$. Such a fit is illustrated in the inset of Fig.~\ref{fig:JJ}a by the dotted line. We chose sections above $|V|=300~\mathrm{\mu V}$ which, together with the current bias range of $\pm 2~\mathrm{\mu A}$, limited us to $V_\mathrm{TG} < -3.07~V$. The results are plotted in purple in Fig.~\ref{fig:JJ}b for both positive and negative $I_\mathrm{exc}$. Denoting the slopes of the high-$|V|$ fits as $R_{Ne}$, we have have calculated the product $I_\mathrm{exc} \cdot R_{Ne}$: it fluctuates between 60--100~$\mathrm{\mu V}$ and is roughly independent of $V_\mathrm{TG}$.


Considering that in other similar top gated samples the mean free path $l$ rarely exceeded 250~nm, this JJ is in the diffusive regime. For diffusive junctions, the superconducting coherence length within InAs can be calculated using $\xi = \sqrt{ \hbar D / \Delta } $ where $D = \sigma / e^2 g(\varepsilon)$ is the diffusion constant\cite{Bonch1990}. For example, at $V_\mathrm{TG} = -3.8~\mathrm{V}$ the normal state conductivity ($\sigma \propto R_{Ne}^{-1}$) is approximately $14.5~\mathrm{mS}$ and, using $m/m_e = 0.03$ in the density of states $g = m / \pi \hbar^2$ of the 1st subband, we get $D \approx 0.72 ~ \mathrm{m^2/s}$ and $\xi \approx 2 ~ \mathrm{\mu m}$, which increases with $V_\mathrm{TG}$. Consequently, the JJ is shorter than $\xi$. In such a device with transparent interfaces, we expect\cite{Shabani2016, Mayer2019} $I_c R_N \approx 1.32 \cdot \pi \Delta / 2 e \approx 259 ~ \mathrm{\mu V}$ and $I_\mathrm{exc} R_{Ne} = \left( \pi^2 / 4 - 1 \right) \Delta / e \approx 183~ \mathrm{\mu V}$ using the value of $\Delta$ calculated from MAR. The experimental values of the two products (30, and 60 to 100~$\mathrm{\mu V}$, respectively) are smaller, which may be caused by less than perfect transmission on the interface. We should note, however, that $R_N$ was always calculated at $I=-1.8~\mathrm{\mu A}$, where the value of $|V|$ is well inside the superconducting gap in almost the full range of $V_\mathrm{TG}$, therefore $R_N$ and $I_c \cdot R_N \approx 30~\mathrm{\mu V}$ are likely underestimated. 


Lastly, we discuss the periodicity of the Fraunhofer pattern of the device at $V_\mathrm{TG}= 2~\mathrm{V}$ as a function of an out of plane magnetic field, as shown in Fig.~\ref{fig:JJ}d. In the simplest approach, one expects nodes at $ B = n \cdot \Delta B_0$ for $n=1,2,3,...$ and spacing $\Delta B_0 = \phi_0 / W L \approx 765~\mathrm{\mu T}$, using the flux quantum $\phi_0 = h/2e = 2.07\cdot 10^{-15}~\mathrm{Wb}$. In contrast, the first node can be found at $B_1 \approx 17~ \mathrm{ \mu T}$, and the subsequent nodes $B_n$ have a slowly increasing spacing $\Delta B$. As a possible explanation, one may consider flux focusing due to the Meissner effect: the magnetic field does not penetrate into the bulk of the wide Al contacts, and is concentrated in and around the JJ, leading to an effective field $B_{eff} > B$ and a node spacing $\Delta B < \Delta B_0 $. Its effect has been estimated in Ref.~\onlinecite{Suominen2017a}, based on the effective out of plane magnetic field around a thin and infinitely long superconducting strip\cite{Zeldov1994}. However, the latter model only applies if $\Lambda \ll W$ where $\Lambda$ is the penetration depth of the magnetic field into the superconductor and $2W$ is the width of the strip. As we shall explain below, this is not the case.

We need to consider two effects. Firstly, due to disorder in the Al layer, the London penetration depth is increased to $\lambda = \lambda_\mathrm{Al} \sqrt{1+\xi_\mathrm{Al}/l_\mathrm{Al}}\approx 203 ~ \mathrm{nm}$. Here we used $\lambda_\mathrm{Al} = 16~\mathrm{nm}$ and $\xi_\mathrm{Al} = 1.6~ \mathrm{\mu m}$ as the London penetration depth and Pippard coherence length\cite{Merservey1969} for bulk Al. $l_\mathrm{Al}$ is the mean free path\cite{Tinkham1996} in the superconductor which we approximated\cite{Suominen2017a} by the Al film thickness $d=10~\mathrm{nm}$. Secondly, in a thin film the effective penetration length\cite{Gubin2005} is $\Lambda = \lambda \coth ( d/\lambda ) $ which, in the thin film limit $d \ll \lambda$, is the Pearl length\cite{Pearl1964} $\Lambda = \lambda^2 / d \approx 4.1~\mathrm{\mu m} $. This is comparable to the size of the Al leads, consequently, the magnetic field penetrates on a large scale into the superconductor, and flux focusing only weakly contributes to the decreased node spacing. We rather attribute it to the effective lengthening of the junction $L_{eff} = L + 2 \Lambda \approx 8.5~ \mathrm{ \mu m}$, from which we expect a node spacing on the order of $27~ \mathrm{\mu T}$, which is comparable to the observed values. We attribute the discrepancy to the fact that $W$ is not much larger than $\Lambda$. As seen in another JJ of $W=4.5~\mathrm{\mu m}$ ($L=0.5~\mathrm{\mu m}$, $B_1=224~\mathrm{\mu T}$, not shown) or in Ref.~\onlinecite{Suominen2017a} of $W=1.5~\mathrm{\mu m}$ ($L=0.45~\mathrm{\mu m}$, $B_1=970~\mathrm{\mu T}$), where $W$ is comparable to or smaller than the Pearl length, this approach gives qualitatively wrong predictions to $B_1$: $53~\mathrm{\mu T}$ and $159~\mathrm{\mu T}$, respectively. The reason is that the screening currents near the edges of the superconducting leads parallel to the applied current direction, with a characteristic width $\Lambda$, can no longer be neglected. In the extreme 2D limit\cite{Baumgartner2021,Rodan2021} $W \ll \Lambda$, the periodicity of short junctions ($L \ll W$) is expected to be proportional to $W^{-2}$[48, 49].


\section{Conclusions}

We have fabricated several Hall bar and Josephson junction devices of a near-surface 2DEG based on InAs. From low-temperature magnetotransport measurements we have calculated that their mobility and mean free path without a dielectric layer are on the order of $10^5 ~ \mathrm{cm^2/Vs}$ and $1~\mathrm{\mu m}$, respectively. This result is relatively high among near-surface 2DEGs, and is independent of the removal of an epitaxial Al layer. We have determined the effective mass and the elastic and transport scattering times. In top gated samples the mobility was consistently reduced and the Dingle temperature increased, indicating increased backscattering and disorder due to the fabrication of the $\mathrm{Al_2 O_3}$ dielectric and the top gate. We have observed conductance quantization in a quantum point contact, and determined the out of plane g-factor: $ \left| g_{\perp}  \right| = 15.8 $.

In a $0.3 ~ \mathrm{\mu m}$ long and $9 ~ \mathrm{\mu m}$ wide Josephson junction we observed a gate-tunable critical current up to $1.5~\mathrm{\mu A}$. We calculated that the induced gap is $\Delta = 125~\mathrm{\mu e V}$, a similar value to Ref.~\onlinecite{Baumgartner2021}. We estimated that the semiconductor-superconductor interface is relatively transparent, and that the junction is in the short diffusive limit. We studied the Fraunhofer pattern in an out of plane magnetic field, and found that the periodicity is reduced due to the effective lengthening of the junction on the order of the Pearl length.

Our results indicate that these InAs 2DEGs are of high quality for the fabrication of quantum point contacts and Josephson junctions with transparent interfaces, and  can serve as a platform for realizing nanostructures in future quantum computing applications.






\begin{acknowledgments}

Author contributions: growth was optimized and carried out by M. K and G. B. Fabrication was done by M. S. Transport measurements and data analysis were carried out by M. S., T. P. and E.T. All authors discussed the results and worked on the manuscript. P. M., S. C. and E.T. proposed the device concept and guided the project.

This work has received funding from the SuperTop QuantERA network, the FET Open AndQC, the FET Open SuperGate, the COST Nanocohybri and FlagERA MultiSpin networks. The authors acknowledge financial support from the Ministry of Innovation and Technology of Hungary, the National Research, Development, and Innovation Office within the Quantum Information National Laboratory of Hungary, the Quantum Technology National Excellence Program (Project Nr. 2017-1.2.1-NKP-2017-00001), as well as Bolyai+ grant UNKP-21-5-BME-343, OTKA grants PD 134758 and K 138433, and the Bolyai Fellowship of the Hungarian Academy of Sciences (Grant BO/00242/20/11). 

We thank F. Nichele and P. Neumann for helpful discussions, and F. F{\"u}l{\"o}p and M. Hajdu for their technical support. The authors are thankful to the Institute of Technical Physics and Materials Science of the Centre for Energy Research for providing their facilities for sample fabrication.

\end{acknowledgments}








\bibliographystyle{apsrev}
\bibliography{main}

\begin{thebibliography}{47}
\expandafter\ifx\csname natexlab\endcsname\relax\def\natexlab#1{#1}\fi
\expandafter\ifx\csname bibnamefont\endcsname\relax
  \def\bibnamefont#1{#1}\fi
\expandafter\ifx\csname bibfnamefont\endcsname\relax
  \def\bibfnamefont#1{#1}\fi
\expandafter\ifx\csname citenamefont\endcsname\relax
  \def\citenamefont#1{#1}\fi
\expandafter\ifx\csname url\endcsname\relax
  \def\url#1{\texttt{#1}}\fi
\expandafter\ifx\csname urlprefix\endcsname\relax\def\urlprefix{URL }\fi
\providecommand{\bibinfo}[2]{#2}
\providecommand{\eprint}[2][]{\url{#2}}

\bibitem[{\citenamefont{Kjaergaard et~al.}(2016)\citenamefont{Kjaergaard,
  Nichele, Suominen, Nowak, Wimmer, Akhmerov, Folk, Flensberg, Shabani,
  Palmstr{\o}m et~al.}}]{Kjaergaard2016a}
\bibinfo{author}{\bibfnamefont{M.}~\bibnamefont{Kjaergaard}},
  \bibinfo{author}{\bibfnamefont{F.}~\bibnamefont{Nichele}},
  \bibinfo{author}{\bibfnamefont{H.~J.} \bibnamefont{Suominen}},
  \bibinfo{author}{\bibfnamefont{M.~P.} \bibnamefont{Nowak}},
  \bibinfo{author}{\bibfnamefont{M.}~\bibnamefont{Wimmer}},
  \bibinfo{author}{\bibfnamefont{A.~R.} \bibnamefont{Akhmerov}},
  \bibinfo{author}{\bibfnamefont{J.~A.} \bibnamefont{Folk}},
  \bibinfo{author}{\bibfnamefont{K.}~\bibnamefont{Flensberg}},
  \bibinfo{author}{\bibfnamefont{J.}~\bibnamefont{Shabani}},
  \bibinfo{author}{\bibfnamefont{C.~J.} \bibnamefont{Palmstr{\o}m}},
  \bibnamefont{et~al.}, \bibinfo{journal}{Nature Communications}
  \textbf{\bibinfo{volume}{7}}, \bibinfo{pages}{1} (\bibinfo{year}{2016}), ISSN
  \bibinfo{issn}{2041-1723}, \eprint{1603.01852},
  \urlprefix\url{https://www.nature.com/articles/ncomms12841}.

\bibitem[{\citenamefont{Suominen
  et~al.}(2017{\natexlab{a}})\citenamefont{Suominen, Kjaergaard, Hamilton,
  Shabani, Palmstr{\o}m, Marcus, and Nichele}}]{Suominen2017}
\bibinfo{author}{\bibfnamefont{H.~J.} \bibnamefont{Suominen}},
  \bibinfo{author}{\bibfnamefont{M.}~\bibnamefont{Kjaergaard}},
  \bibinfo{author}{\bibfnamefont{A.~R.} \bibnamefont{Hamilton}},
  \bibinfo{author}{\bibfnamefont{J.}~\bibnamefont{Shabani}},
  \bibinfo{author}{\bibfnamefont{C.~J.} \bibnamefont{Palmstr{\o}m}},
  \bibinfo{author}{\bibfnamefont{C.~M.} \bibnamefont{Marcus}},
  \bibnamefont{and} \bibinfo{author}{\bibfnamefont{F.}~\bibnamefont{Nichele}},
  \bibinfo{journal}{Physical Review Letters} \textbf{\bibinfo{volume}{119}},
  \bibinfo{pages}{176805} (\bibinfo{year}{2017}{\natexlab{a}}), ISSN
  \bibinfo{issn}{10797114}, \eprint{1703.03699}.

\bibitem[{\citenamefont{Moehle et~al.}(2021)\citenamefont{Moehle, Ke, Wang,
  Thomas, Xiao, Karwal, Lodari, {Van De Kerkhof}, Termaat, Gardner
  et~al.}}]{Moehle2021}
\bibinfo{author}{\bibfnamefont{C.~M.} \bibnamefont{Moehle}},
  \bibinfo{author}{\bibfnamefont{C.~T.} \bibnamefont{Ke}},
  \bibinfo{author}{\bibfnamefont{Q.}~\bibnamefont{Wang}},
  \bibinfo{author}{\bibfnamefont{C.}~\bibnamefont{Thomas}},
  \bibinfo{author}{\bibfnamefont{D.}~\bibnamefont{Xiao}},
  \bibinfo{author}{\bibfnamefont{S.}~\bibnamefont{Karwal}},
  \bibinfo{author}{\bibfnamefont{M.}~\bibnamefont{Lodari}},
  \bibinfo{author}{\bibfnamefont{V.}~\bibnamefont{{Van De Kerkhof}}},
  \bibinfo{author}{\bibfnamefont{R.}~\bibnamefont{Termaat}},
  \bibinfo{author}{\bibfnamefont{G.~C.} \bibnamefont{Gardner}},
  \bibnamefont{et~al.}, \bibinfo{journal}{Nano Letters}
  \textbf{\bibinfo{volume}{21}}, \bibinfo{pages}{9990} (\bibinfo{year}{2021}),
  ISSN \bibinfo{issn}{15306992}, \eprint{2105.10437},
  \urlprefix\url{https://pubs.acs.org/doi/abs/10.1021/acs.nanolett.1c03520}.

\bibitem[{\citenamefont{Kitaev}(2001)}]{Kitaev2001}
\bibinfo{author}{\bibfnamefont{A.~Y.} \bibnamefont{Kitaev}},
  \bibinfo{journal}{Physics-Uspekhi} \textbf{\bibinfo{volume}{44}},
  \bibinfo{pages}{131} (\bibinfo{year}{2001}), ISSN \bibinfo{issn}{1063-7869},
  \eprint{0010440},
  \urlprefix\url{https://iopscience.iop.org/article/10.1070/1063-7869/44/10S/S29
  https://iopscience.iop.org/article/10.1070/1063-7869/44/10S/S29/meta}.

\bibitem[{\citenamefont{Oreg et~al.}(2010)\citenamefont{Oreg, Refael, and von
  Oppen}}]{Oreg2010}
\bibinfo{author}{\bibfnamefont{Y.}~\bibnamefont{Oreg}},
  \bibinfo{author}{\bibfnamefont{G.}~\bibnamefont{Refael}}, \bibnamefont{and}
  \bibinfo{author}{\bibfnamefont{F.}~\bibnamefont{von Oppen}},
  \bibinfo{journal}{Physical Review Letters} \textbf{\bibinfo{volume}{105}},
  \bibinfo{pages}{177002} (\bibinfo{year}{2010}), ISSN
  \bibinfo{issn}{00319007}, \eprint{1003.1145},
  \urlprefix\url{https://journals.aps.org/prl/abstract/10.1103/PhysRevLett.105.177002}.

\bibitem[{\citenamefont{Lutchyn et~al.}(2010)\citenamefont{Lutchyn, Sau, and
  {Das Sarma}}}]{Lutchyn2010}
\bibinfo{author}{\bibfnamefont{R.~M.} \bibnamefont{Lutchyn}},
  \bibinfo{author}{\bibfnamefont{J.~D.} \bibnamefont{Sau}}, \bibnamefont{and}
  \bibinfo{author}{\bibfnamefont{S.}~\bibnamefont{{Das Sarma}}},
  \bibinfo{journal}{Physical Review Letters} \textbf{\bibinfo{volume}{105}},
  \bibinfo{pages}{077001} (\bibinfo{year}{2010}), ISSN
  \bibinfo{issn}{00319007}, \eprint{1002.4033},
  \urlprefix\url{https://journals.aps.org/prl/abstract/10.1103/PhysRevLett.105.077001}.

\bibitem[{\citenamefont{Casparis et~al.}(2018)\citenamefont{Casparis, Connolly,
  Kjaergaard, Pearson, Kringh{\o}j, Larsen, Kuemmeth, Wang, Thomas, Gronin
  et~al.}}]{Casparis2018}
\bibinfo{author}{\bibfnamefont{L.}~\bibnamefont{Casparis}},
  \bibinfo{author}{\bibfnamefont{M.~R.} \bibnamefont{Connolly}},
  \bibinfo{author}{\bibfnamefont{M.}~\bibnamefont{Kjaergaard}},
  \bibinfo{author}{\bibfnamefont{N.~J.} \bibnamefont{Pearson}},
  \bibinfo{author}{\bibfnamefont{A.}~\bibnamefont{Kringh{\o}j}},
  \bibinfo{author}{\bibfnamefont{T.~W.} \bibnamefont{Larsen}},
  \bibinfo{author}{\bibfnamefont{F.}~\bibnamefont{Kuemmeth}},
  \bibinfo{author}{\bibfnamefont{T.}~\bibnamefont{Wang}},
  \bibinfo{author}{\bibfnamefont{C.}~\bibnamefont{Thomas}},
  \bibinfo{author}{\bibfnamefont{S.}~\bibnamefont{Gronin}},
  \bibnamefont{et~al.}, \emph{\bibinfo{title}{{Superconducting gatemon qubit
  based on a proximitized two-dimensional electron gas}}}
  (\bibinfo{year}{2018}), \eprint{1711.07665}.

\bibitem[{\citenamefont{Albrecht et~al.}(2016)\citenamefont{Albrecht,
  Higginbotham, Madsen, Kuemmeth, Jespersen, Nyg{\aa}rd, Krogstrup, and
  Marcus}}]{Albrecht2016}
\bibinfo{author}{\bibfnamefont{S.~M.} \bibnamefont{Albrecht}},
  \bibinfo{author}{\bibfnamefont{A.~P.} \bibnamefont{Higginbotham}},
  \bibinfo{author}{\bibfnamefont{M.}~\bibnamefont{Madsen}},
  \bibinfo{author}{\bibfnamefont{F.}~\bibnamefont{Kuemmeth}},
  \bibinfo{author}{\bibfnamefont{T.~S.} \bibnamefont{Jespersen}},
  \bibinfo{author}{\bibfnamefont{J.}~\bibnamefont{Nyg{\aa}rd}},
  \bibinfo{author}{\bibfnamefont{P.}~\bibnamefont{Krogstrup}},
  \bibnamefont{and} \bibinfo{author}{\bibfnamefont{C.~M.}
  \bibnamefont{Marcus}}, \bibinfo{journal}{Nature}
  \textbf{\bibinfo{volume}{531}}, \bibinfo{pages}{206} (\bibinfo{year}{2016}),
  ISSN \bibinfo{issn}{1476-4687}, \eprint{1603.03217},
  \urlprefix\url{https://www.nature.com/articles/nature17162}.

\bibitem[{\citenamefont{Fornieri et~al.}(2019)\citenamefont{Fornieri, Whiticar,
  Setiawan, Portol{\'{e}}s, Drachmann, Keselman, Gronin, Thomas, Wang, Kallaher
  et~al.}}]{Fornieri2019}
\bibinfo{author}{\bibfnamefont{A.}~\bibnamefont{Fornieri}},
  \bibinfo{author}{\bibfnamefont{A.~M.} \bibnamefont{Whiticar}},
  \bibinfo{author}{\bibfnamefont{F.}~\bibnamefont{Setiawan}},
  \bibinfo{author}{\bibfnamefont{E.}~\bibnamefont{Portol{\'{e}}s}},
  \bibinfo{author}{\bibfnamefont{A.~C.} \bibnamefont{Drachmann}},
  \bibinfo{author}{\bibfnamefont{A.}~\bibnamefont{Keselman}},
  \bibinfo{author}{\bibfnamefont{S.}~\bibnamefont{Gronin}},
  \bibinfo{author}{\bibfnamefont{C.}~\bibnamefont{Thomas}},
  \bibinfo{author}{\bibfnamefont{T.}~\bibnamefont{Wang}},
  \bibinfo{author}{\bibfnamefont{R.}~\bibnamefont{Kallaher}},
  \bibnamefont{et~al.}, \emph{\bibinfo{title}{{Evidence of topological
  superconductivity in planar Josephson junctions}}} (\bibinfo{year}{2019}),
  \eprint{1809.03037}.

\bibitem[{\citenamefont{Dartiailh et~al.}(2021)\citenamefont{Dartiailh, Mayer,
  Yuan, Wickramasinghe, Matos-Abiague, {\v{Z}}uti{\'{c}}, and
  Shabani}}]{Dartiailh2021}
\bibinfo{author}{\bibfnamefont{M.~C.} \bibnamefont{Dartiailh}},
  \bibinfo{author}{\bibfnamefont{W.}~\bibnamefont{Mayer}},
  \bibinfo{author}{\bibfnamefont{J.}~\bibnamefont{Yuan}},
  \bibinfo{author}{\bibfnamefont{K.~S.} \bibnamefont{Wickramasinghe}},
  \bibinfo{author}{\bibfnamefont{A.}~\bibnamefont{Matos-Abiague}},
  \bibinfo{author}{\bibfnamefont{I.}~\bibnamefont{{\v{Z}}uti{\'{c}}}},
  \bibnamefont{and} \bibinfo{author}{\bibfnamefont{J.}~\bibnamefont{Shabani}},
  \bibinfo{journal}{Physical Review Letters} \textbf{\bibinfo{volume}{126}},
  \bibinfo{pages}{036802} (\bibinfo{year}{2021}), ISSN
  \bibinfo{issn}{10797114}, \eprint{1906.01179},
  \urlprefix\url{https://journals.aps.org/prl/abstract/10.1103/PhysRevLett.126.036802}.

\bibitem[{\citenamefont{Tschirky et~al.}(2017)\citenamefont{Tschirky, Mueller,
  Lehner, F{\"{a}}lt, Ihn, Ensslin, and Wegscheider}}]{Tschirky2017}
\bibinfo{author}{\bibfnamefont{T.}~\bibnamefont{Tschirky}},
  \bibinfo{author}{\bibfnamefont{S.}~\bibnamefont{Mueller}},
  \bibinfo{author}{\bibfnamefont{C.~A.} \bibnamefont{Lehner}},
  \bibinfo{author}{\bibfnamefont{S.}~\bibnamefont{F{\"{a}}lt}},
  \bibinfo{author}{\bibfnamefont{T.}~\bibnamefont{Ihn}},
  \bibinfo{author}{\bibfnamefont{K.}~\bibnamefont{Ensslin}}, \bibnamefont{and}
  \bibinfo{author}{\bibfnamefont{W.}~\bibnamefont{Wegscheider}},
  \bibinfo{journal}{Physical Review B} \textbf{\bibinfo{volume}{95}},
  \bibinfo{pages}{115304} (\bibinfo{year}{2017}), ISSN
  \bibinfo{issn}{24699969}, \eprint{1612.06782},
  \urlprefix\url{https://journals.aps.org/prb/abstract/10.1103/PhysRevB.95.115304}.

\bibitem[{\citenamefont{Yuan et~al.}(2020)\citenamefont{Yuan, Hatefipour,
  Magill, Mayer, Dartiailh, Sardashti, Wickramasinghe, Khodaparast, Matsuda,
  Kohama et~al.}}]{Yuan2020}
\bibinfo{author}{\bibfnamefont{J.}~\bibnamefont{Yuan}},
  \bibinfo{author}{\bibfnamefont{M.}~\bibnamefont{Hatefipour}},
  \bibinfo{author}{\bibfnamefont{B.~A.} \bibnamefont{Magill}},
  \bibinfo{author}{\bibfnamefont{W.}~\bibnamefont{Mayer}},
  \bibinfo{author}{\bibfnamefont{M.~C.} \bibnamefont{Dartiailh}},
  \bibinfo{author}{\bibfnamefont{K.}~\bibnamefont{Sardashti}},
  \bibinfo{author}{\bibfnamefont{K.~S.} \bibnamefont{Wickramasinghe}},
  \bibinfo{author}{\bibfnamefont{G.~A.} \bibnamefont{Khodaparast}},
  \bibinfo{author}{\bibfnamefont{Y.~H.} \bibnamefont{Matsuda}},
  \bibinfo{author}{\bibfnamefont{Y.}~\bibnamefont{Kohama}},
  \bibnamefont{et~al.}, \bibinfo{journal}{Physical Review B}
  \textbf{\bibinfo{volume}{101}}, \bibinfo{pages}{205310}
  (\bibinfo{year}{2020}), ISSN \bibinfo{issn}{24699969}, \eprint{1911.02738},
  \urlprefix\url{https://journals.aps.org/prb/abstract/10.1103/PhysRevB.101.205310}.

\bibitem[{\citenamefont{Wickramasinghe
  et~al.}(2018)\citenamefont{Wickramasinghe, Mayer, Yuan, Nguyen, Jiao,
  Manucharyan, and Shabani}}]{Wickramasinghe2018}
\bibinfo{author}{\bibfnamefont{K.~S.} \bibnamefont{Wickramasinghe}},
  \bibinfo{author}{\bibfnamefont{W.}~\bibnamefont{Mayer}},
  \bibinfo{author}{\bibfnamefont{J.}~\bibnamefont{Yuan}},
  \bibinfo{author}{\bibfnamefont{T.}~\bibnamefont{Nguyen}},
  \bibinfo{author}{\bibfnamefont{L.}~\bibnamefont{Jiao}},
  \bibinfo{author}{\bibfnamefont{V.}~\bibnamefont{Manucharyan}},
  \bibnamefont{and} \bibinfo{author}{\bibfnamefont{J.}~\bibnamefont{Shabani}},
  \bibinfo{journal}{Applied Physics Letters} \textbf{\bibinfo{volume}{113}},
  \bibinfo{pages}{262104} (\bibinfo{year}{2018}), ISSN
  \bibinfo{issn}{0003-6951}, \eprint{1802.09569},
  \urlprefix\url{https://aip.scitation.org/doi/abs/10.1063/1.5050413}.

\bibitem[{\citenamefont{Capotondi
  et~al.}(2005{\natexlab{a}})\citenamefont{Capotondi, Biasiol, Ercolani, and
  Sorba}}]{Capotondi2005a}
\bibinfo{author}{\bibfnamefont{F.}~\bibnamefont{Capotondi}},
  \bibinfo{author}{\bibfnamefont{G.}~\bibnamefont{Biasiol}},
  \bibinfo{author}{\bibfnamefont{D.}~\bibnamefont{Ercolani}}, \bibnamefont{and}
  \bibinfo{author}{\bibfnamefont{L.}~\bibnamefont{Sorba}},
  \bibinfo{journal}{Journal of Crystal Growth} \textbf{\bibinfo{volume}{278}},
  \bibinfo{pages}{538} (\bibinfo{year}{2005}{\natexlab{a}}), ISSN
  \bibinfo{issn}{0022-0248}.

\bibitem[{\citenamefont{Ercolani et~al.}(2008)\citenamefont{Ercolani, Biasiol,
  Cancellieri, Rosini, Jacoboni, Carillo, Heun, Sorba, and
  Nolting}}]{Ercolani2008}
\bibinfo{author}{\bibfnamefont{D.}~\bibnamefont{Ercolani}},
  \bibinfo{author}{\bibfnamefont{G.}~\bibnamefont{Biasiol}},
  \bibinfo{author}{\bibfnamefont{E.}~\bibnamefont{Cancellieri}},
  \bibinfo{author}{\bibfnamefont{M.}~\bibnamefont{Rosini}},
  \bibinfo{author}{\bibfnamefont{C.}~\bibnamefont{Jacoboni}},
  \bibinfo{author}{\bibfnamefont{F.}~\bibnamefont{Carillo}},
  \bibinfo{author}{\bibfnamefont{S.}~\bibnamefont{Heun}},
  \bibinfo{author}{\bibfnamefont{L.}~\bibnamefont{Sorba}}, \bibnamefont{and}
  \bibinfo{author}{\bibfnamefont{F.}~\bibnamefont{Nolting}},
  \bibinfo{journal}{Physical Review B - Condensed Matter and Materials Physics}
  \textbf{\bibinfo{volume}{77}}, \bibinfo{pages}{235307}
  (\bibinfo{year}{2008}), ISSN \bibinfo{issn}{10980121},
  \urlprefix\url{https://journals.aps.org/prb/abstract/10.1103/PhysRevB.77.235307}.

\bibitem[{\citenamefont{Hatke et~al.}(2017)\citenamefont{Hatke, Wang, Thomas,
  Gardner, and Manfra}}]{Hatke2017}
\bibinfo{author}{\bibfnamefont{A.~T.} \bibnamefont{Hatke}},
  \bibinfo{author}{\bibfnamefont{T.}~\bibnamefont{Wang}},
  \bibinfo{author}{\bibfnamefont{C.}~\bibnamefont{Thomas}},
  \bibinfo{author}{\bibfnamefont{G.~C.} \bibnamefont{Gardner}},
  \bibnamefont{and} \bibinfo{author}{\bibfnamefont{M.~J.}
  \bibnamefont{Manfra}}, \bibinfo{journal}{Applied Physics Letters}
  \textbf{\bibinfo{volume}{111}}, \bibinfo{pages}{142106}
  (\bibinfo{year}{2017}), ISSN \bibinfo{issn}{0003-6951}, \eprint{1707.00031},
  \urlprefix\url{https://aip.scitation.org/doi/abs/10.1063/1.4993784}.

\bibitem[{\citenamefont{Capotondi et~al.}(2004)\citenamefont{Capotondi,
  Biasiol, Vobornik, Sorba, Giazotto, Cavallini, and Fraboni}}]{Capotondi2004}
\bibinfo{author}{\bibfnamefont{F.}~\bibnamefont{Capotondi}},
  \bibinfo{author}{\bibfnamefont{G.}~\bibnamefont{Biasiol}},
  \bibinfo{author}{\bibfnamefont{I.}~\bibnamefont{Vobornik}},
  \bibinfo{author}{\bibfnamefont{L.}~\bibnamefont{Sorba}},
  \bibinfo{author}{\bibfnamefont{F.}~\bibnamefont{Giazotto}},
  \bibinfo{author}{\bibfnamefont{A.}~\bibnamefont{Cavallini}},
  \bibnamefont{and} \bibinfo{author}{\bibfnamefont{B.}~\bibnamefont{Fraboni}},
  \bibinfo{journal}{Journal of Vacuum Science \& Technology B: Microelectronics
  and Nanometer Structures Processing, Measurement, and Phenomena}
  \textbf{\bibinfo{volume}{22}}, \bibinfo{pages}{702} (\bibinfo{year}{2004}),
  ISSN \bibinfo{issn}{1071-1023},
  \urlprefix\url{https://avs.scitation.org/doi/abs/10.1116/1.1688345}.

\bibitem[{\citenamefont{Benali et~al.}(2022)\citenamefont{Benali, Rajak,
  Ciancio, Plaisier, Heun, and Biasiol}}]{Benali2022}
\bibinfo{author}{\bibfnamefont{A.}~\bibnamefont{Benali}},
  \bibinfo{author}{\bibfnamefont{P.}~\bibnamefont{Rajak}},
  \bibinfo{author}{\bibfnamefont{R.}~\bibnamefont{Ciancio}},
  \bibinfo{author}{\bibfnamefont{J.~R.} \bibnamefont{Plaisier}},
  \bibinfo{author}{\bibfnamefont{S.}~\bibnamefont{Heun}}, \bibnamefont{and}
  \bibinfo{author}{\bibfnamefont{G.}~\bibnamefont{Biasiol}},
  \bibinfo{journal}{Journal of Crystal Growth} \textbf{\bibinfo{volume}{593}},
  \bibinfo{pages}{126768} (\bibinfo{year}{2022}),
  \urlprefix\url{https://doi.org/10.1016/j.jcrysgro.2022.126768}.

\bibitem[{\citenamefont{Lee et~al.}(2019)\citenamefont{Lee, Shojaei,
  Pendharkar, McFadden, Kim, Suominen, Kjaergaard, Nichele, Zhang, Marcus
  et~al.}}]{Lee2019}
\bibinfo{author}{\bibfnamefont{J.~S.} \bibnamefont{Lee}},
  \bibinfo{author}{\bibfnamefont{B.}~\bibnamefont{Shojaei}},
  \bibinfo{author}{\bibfnamefont{M.}~\bibnamefont{Pendharkar}},
  \bibinfo{author}{\bibfnamefont{A.~P.} \bibnamefont{McFadden}},
  \bibinfo{author}{\bibfnamefont{Y.}~\bibnamefont{Kim}},
  \bibinfo{author}{\bibfnamefont{H.~J.} \bibnamefont{Suominen}},
  \bibinfo{author}{\bibfnamefont{M.}~\bibnamefont{Kjaergaard}},
  \bibinfo{author}{\bibfnamefont{F.}~\bibnamefont{Nichele}},
  \bibinfo{author}{\bibfnamefont{H.}~\bibnamefont{Zhang}},
  \bibinfo{author}{\bibfnamefont{C.~M.} \bibnamefont{Marcus}},
  \bibnamefont{et~al.}, \bibinfo{journal}{Nano Letters}
  \textbf{\bibinfo{volume}{19}}, \bibinfo{pages}{3083} (\bibinfo{year}{2019}),
  ISSN \bibinfo{issn}{15306992},
  \urlprefix\url{https://pubs.acs.org/doi/abs/10.1021/acs.nanolett.9b00494}.

\bibitem[{\citenamefont{Mayer et~al.}(2019)\citenamefont{Mayer, Yuan,
  Wickramasinghe, Nguyen, Dartiailh, and Shabani}}]{Mayer2019}
\bibinfo{author}{\bibfnamefont{W.}~\bibnamefont{Mayer}},
  \bibinfo{author}{\bibfnamefont{J.}~\bibnamefont{Yuan}},
  \bibinfo{author}{\bibfnamefont{K.~S.} \bibnamefont{Wickramasinghe}},
  \bibinfo{author}{\bibfnamefont{T.}~\bibnamefont{Nguyen}},
  \bibinfo{author}{\bibfnamefont{M.~C.} \bibnamefont{Dartiailh}},
  \bibnamefont{and} \bibinfo{author}{\bibfnamefont{J.}~\bibnamefont{Shabani}},
  \bibinfo{journal}{Applied Physics Letters} \textbf{\bibinfo{volume}{114}}
  (\bibinfo{year}{2019}), ISSN \bibinfo{issn}{00036951}, \eprint{1810.02514}.

\bibitem[{\citenamefont{Yuan et~al.}(2021)\citenamefont{Yuan, Wickramasinghe,
  Strickland, Dartiailh, Sardashti, Hatefipour, and Shabani}}]{Yuan2021}
\bibinfo{author}{\bibfnamefont{J.~O.} \bibnamefont{Yuan}},
  \bibinfo{author}{\bibfnamefont{K.~S.} \bibnamefont{Wickramasinghe}},
  \bibinfo{author}{\bibfnamefont{W.~M.} \bibnamefont{Strickland}},
  \bibinfo{author}{\bibfnamefont{M.~C.} \bibnamefont{Dartiailh}},
  \bibinfo{author}{\bibfnamefont{K.}~\bibnamefont{Sardashti}},
  \bibinfo{author}{\bibfnamefont{M.}~\bibnamefont{Hatefipour}},
  \bibnamefont{and} \bibinfo{author}{\bibfnamefont{J.}~\bibnamefont{Shabani}},
  \bibinfo{journal}{Journal of Vacuum Science \& Technology A: Vacuum,
  Surfaces, and Films} \textbf{\bibinfo{volume}{39}}, \bibinfo{pages}{033407}
  (\bibinfo{year}{2021}), ISSN \bibinfo{issn}{0734-2101}, \eprint{2104.01159},
  \urlprefix\url{https://avs.scitation.org/doi/abs/10.1116/6.0000918}.

\bibitem[{\citenamefont{Tan et~al.}(1990)\citenamefont{Tan, Snider, Chang, and
  Hu}}]{Tan1990}
\bibinfo{author}{\bibfnamefont{I.~H.} \bibnamefont{Tan}},
  \bibinfo{author}{\bibfnamefont{G.~L.} \bibnamefont{Snider}},
  \bibinfo{author}{\bibfnamefont{L.~D.} \bibnamefont{Chang}}, \bibnamefont{and}
  \bibinfo{author}{\bibfnamefont{E.~L.} \bibnamefont{Hu}},
  \bibinfo{journal}{Journal of Applied Physics} \textbf{\bibinfo{volume}{68}},
  \bibinfo{pages}{4071} (\bibinfo{year}{1990}), ISSN \bibinfo{issn}{0021-8979},
  \urlprefix\url{https://aip.scitation.org/doi/abs/10.1063/1.346245}.

\bibitem[{\citenamefont{Capotondi
  et~al.}(2005{\natexlab{b}})\citenamefont{Capotondi, Biasiol, Ercolani,
  Grillo, Carlino, Romanato, and Sorba}}]{Capotondi2005}
\bibinfo{author}{\bibfnamefont{F.}~\bibnamefont{Capotondi}},
  \bibinfo{author}{\bibfnamefont{G.}~\bibnamefont{Biasiol}},
  \bibinfo{author}{\bibfnamefont{D.}~\bibnamefont{Ercolani}},
  \bibinfo{author}{\bibfnamefont{V.}~\bibnamefont{Grillo}},
  \bibinfo{author}{\bibfnamefont{E.}~\bibnamefont{Carlino}},
  \bibinfo{author}{\bibfnamefont{F.}~\bibnamefont{Romanato}}, \bibnamefont{and}
  \bibinfo{author}{\bibfnamefont{L.}~\bibnamefont{Sorba}},
  \bibinfo{journal}{Thin Solid Films} \textbf{\bibinfo{volume}{484}},
  \bibinfo{pages}{400} (\bibinfo{year}{2005}{\natexlab{b}}), ISSN
  \bibinfo{issn}{0040-6090}.

\bibitem[{\citenamefont{Kjaergaard}(2015)}]{Kjaergaard2015}
\bibinfo{author}{\bibfnamefont{M.}~\bibnamefont{Kjaergaard}}, Ph.D. thesis,
  \bibinfo{school}{Copenhagen University} (\bibinfo{year}{2015}),
  \urlprefix\url{https://qdev.nbi.ku.dk/student_theses/phdthesis_webversion.pdf}.

\bibitem[{\citenamefont{Shabani et~al.}(2016)\citenamefont{Shabani, Kjaergaard,
  Suominen, Kim, Nichele, Pakrouski, Stankevic, Lutchyn, Krogstrup,
  Feidenhans'L et~al.}}]{Shabani2016}
\bibinfo{author}{\bibfnamefont{J.}~\bibnamefont{Shabani}},
  \bibinfo{author}{\bibfnamefont{M.}~\bibnamefont{Kjaergaard}},
  \bibinfo{author}{\bibfnamefont{H.~J.} \bibnamefont{Suominen}},
  \bibinfo{author}{\bibfnamefont{Y.}~\bibnamefont{Kim}},
  \bibinfo{author}{\bibfnamefont{F.}~\bibnamefont{Nichele}},
  \bibinfo{author}{\bibfnamefont{K.}~\bibnamefont{Pakrouski}},
  \bibinfo{author}{\bibfnamefont{T.}~\bibnamefont{Stankevic}},
  \bibinfo{author}{\bibfnamefont{R.~M.} \bibnamefont{Lutchyn}},
  \bibinfo{author}{\bibfnamefont{P.}~\bibnamefont{Krogstrup}},
  \bibinfo{author}{\bibfnamefont{R.}~\bibnamefont{Feidenhans'L}},
  \bibnamefont{et~al.}, \bibinfo{journal}{Physical Review B}
  \textbf{\bibinfo{volume}{93}}, \bibinfo{pages}{155402}
  (\bibinfo{year}{2016}), ISSN \bibinfo{issn}{24699969}, \eprint{1511.01127}.

\bibitem[{\citenamefont{Shoenberg}(2009)}]{Shoenberg2009}
\bibinfo{author}{\bibfnamefont{D.~D.} \bibnamefont{Shoenberg}},
  \emph{\bibinfo{title}{{Magnetic oscillations in metals}}}
  (\bibinfo{publisher}{Cambridge University Press}, \bibinfo{year}{2009}), ISBN
  \bibinfo{isbn}{978-0521118781}.

\bibitem[{\citenamefont{Mikhailova}(1996)}]{Mikhailova1996}
\bibinfo{author}{\bibfnamefont{M.~P.} \bibnamefont{Mikhailova}}, in
  \emph{\bibinfo{booktitle}{Handbook Series on Semiconductor Parameters}},
  edited by \bibinfo{editor}{\bibfnamefont{M.}~\bibnamefont{Levinshtein}},
  \bibinfo{editor}{\bibfnamefont{S.}~\bibnamefont{Rumyantsev}},
  \bibnamefont{and} \bibinfo{editor}{\bibfnamefont{M.}~\bibnamefont{Shur}}
  (\bibinfo{publisher}{WORLD SCIENTIFIC}, \bibinfo{year}{1996}),
  vol.~\bibinfo{volume}{1}, ISBN \bibinfo{isbn}{978-981-02-2934-4},
  \urlprefix\url{https://www.worldscientific.com/worldscibooks/10.1142/2046-vol1}.

\bibitem[{\citenamefont{Lin-Chung and Yang}(1993)}]{Lin-Chung1993}
\bibinfo{author}{\bibfnamefont{P.~J.} \bibnamefont{Lin-Chung}}
  \bibnamefont{and} \bibinfo{author}{\bibfnamefont{M.~J.} \bibnamefont{Yang}},
  \bibinfo{journal}{Physical Review B} \textbf{\bibinfo{volume}{48}},
  \bibinfo{pages}{5338} (\bibinfo{year}{1993}), ISSN \bibinfo{issn}{01631829},
  \urlprefix\url{https://journals.aps.org/prb/abstract/10.1103/PhysRevB.48.5338}.

\bibitem[{\citenamefont{Monteverde et~al.}(2010)\citenamefont{Monteverde,
  Ojeda-Aristizabal, Weil, Bennaceur, Ferrier, Gu{\'{e}}ron, Glattli, Bouchiat,
  Fuchs, and Maslov}}]{Monteverde2010}
\bibinfo{author}{\bibfnamefont{M.}~\bibnamefont{Monteverde}},
  \bibinfo{author}{\bibfnamefont{C.}~\bibnamefont{Ojeda-Aristizabal}},
  \bibinfo{author}{\bibfnamefont{R.}~\bibnamefont{Weil}},
  \bibinfo{author}{\bibfnamefont{K.}~\bibnamefont{Bennaceur}},
  \bibinfo{author}{\bibfnamefont{M.}~\bibnamefont{Ferrier}},
  \bibinfo{author}{\bibfnamefont{S.}~\bibnamefont{Gu{\'{e}}ron}},
  \bibinfo{author}{\bibfnamefont{C.}~\bibnamefont{Glattli}},
  \bibinfo{author}{\bibfnamefont{H.}~\bibnamefont{Bouchiat}},
  \bibinfo{author}{\bibfnamefont{J.~N.} \bibnamefont{Fuchs}}, \bibnamefont{and}
  \bibinfo{author}{\bibfnamefont{D.~L.} \bibnamefont{Maslov}},
  \bibinfo{journal}{Physical Review Letters} \textbf{\bibinfo{volume}{104}},
  \bibinfo{pages}{126801} (\bibinfo{year}{2010}), ISSN
  \bibinfo{issn}{00319007}, \eprint{0903.3285}.

\bibitem[{\citenamefont{Ellenberger et~al.}(2006)\citenamefont{Ellenberger,
  Simovi{\v{c}}, Leturcq, Ihn, Ulloa, Ensslin, Driscoll, and
  Gossard}}]{Ellenberger2006}
\bibinfo{author}{\bibfnamefont{C.}~\bibnamefont{Ellenberger}},
  \bibinfo{author}{\bibfnamefont{B.}~\bibnamefont{Simovi{\v{c}}}},
  \bibinfo{author}{\bibfnamefont{R.}~\bibnamefont{Leturcq}},
  \bibinfo{author}{\bibfnamefont{T.}~\bibnamefont{Ihn}},
  \bibinfo{author}{\bibfnamefont{S.~E.} \bibnamefont{Ulloa}},
  \bibinfo{author}{\bibfnamefont{K.}~\bibnamefont{Ensslin}},
  \bibinfo{author}{\bibfnamefont{D.~C.} \bibnamefont{Driscoll}},
  \bibnamefont{and} \bibinfo{author}{\bibfnamefont{A.~C.}
  \bibnamefont{Gossard}}, \bibinfo{journal}{Physical Review B - Condensed
  Matter and Materials Physics} \textbf{\bibinfo{volume}{74}},
  \bibinfo{pages}{195313} (\bibinfo{year}{2006}), ISSN
  \bibinfo{issn}{10980121}, \eprint{0602271},
  \urlprefix\url{https://journals.aps.org/prb/abstract/10.1103/PhysRevB.74.195313}.

\bibitem[{\citenamefont{Pauka et~al.}(2020)\citenamefont{Pauka, Witt, Allen,
  Harlech-Jones, Jouan, Gardner, Gronin, Wang, Thomas, Manfra
  et~al.}}]{Pauka2020}
\bibinfo{author}{\bibfnamefont{S.~J.} \bibnamefont{Pauka}},
  \bibinfo{author}{\bibfnamefont{J.~D.} \bibnamefont{Witt}},
  \bibinfo{author}{\bibfnamefont{C.~N.} \bibnamefont{Allen}},
  \bibinfo{author}{\bibfnamefont{B.}~\bibnamefont{Harlech-Jones}},
  \bibinfo{author}{\bibfnamefont{A.}~\bibnamefont{Jouan}},
  \bibinfo{author}{\bibfnamefont{G.~C.} \bibnamefont{Gardner}},
  \bibinfo{author}{\bibfnamefont{S.}~\bibnamefont{Gronin}},
  \bibinfo{author}{\bibfnamefont{T.}~\bibnamefont{Wang}},
  \bibinfo{author}{\bibfnamefont{C.}~\bibnamefont{Thomas}},
  \bibinfo{author}{\bibfnamefont{M.~J.} \bibnamefont{Manfra}},
  \bibnamefont{et~al.}, \bibinfo{journal}{Journal of Applied Physics}
  \textbf{\bibinfo{volume}{128}}, \bibinfo{pages}{114301}
  (\bibinfo{year}{2020}), ISSN \bibinfo{issn}{0021-8979}, \eprint{1908.08689},
  \urlprefix\url{https://aip.scitation.org/doi/abs/10.1063/5.0014361}.

\bibitem[{\citenamefont{Ensslin et~al.}(1993)\citenamefont{Ensslin, Wixforth,
  Sundaram, Hopkins, English, and Gossard}}]{Ensslin1993}
\bibinfo{author}{\bibfnamefont{K.}~\bibnamefont{Ensslin}},
  \bibinfo{author}{\bibfnamefont{A.}~\bibnamefont{Wixforth}},
  \bibinfo{author}{\bibfnamefont{M.}~\bibnamefont{Sundaram}},
  \bibinfo{author}{\bibfnamefont{P.~F.} \bibnamefont{Hopkins}},
  \bibinfo{author}{\bibfnamefont{J.~H.} \bibnamefont{English}},
  \bibnamefont{and} \bibinfo{author}{\bibfnamefont{A.~C.}
  \bibnamefont{Gossard}}, \bibinfo{journal}{Physical Review B}
  \textbf{\bibinfo{volume}{47}}, \bibinfo{pages}{1366} (\bibinfo{year}{1993}),
  ISSN \bibinfo{issn}{01631829},
  \urlprefix\url{https://journals.aps.org/prb/abstract/10.1103/PhysRevB.47.1366}.

\bibitem[{\citenamefont{St{\"{o}}rmer and Gossard}(1982)}]{Stormer1982}
\bibinfo{author}{\bibfnamefont{H.~L.} \bibnamefont{St{\"{o}}rmer}}
  \bibnamefont{and} \bibinfo{author}{\bibfnamefont{A.~C.}
  \bibnamefont{Gossard}}, \bibinfo{journal}{Solid State Communications}
  \textbf{\bibinfo{volume}{41}}, \bibinfo{pages}{707} (\bibinfo{year}{1982}).

\bibitem[{\citenamefont{Pidgeon et~al.}(1967)\citenamefont{Pidgeon, Mitchell,
  and Brown}}]{Pidgeon1967}
\bibinfo{author}{\bibfnamefont{C.~R.} \bibnamefont{Pidgeon}},
  \bibinfo{author}{\bibfnamefont{D.~L.} \bibnamefont{Mitchell}},
  \bibnamefont{and} \bibinfo{author}{\bibfnamefont{R.~N.} \bibnamefont{Brown}},
  \bibinfo{journal}{Physical Review} \textbf{\bibinfo{volume}{154}},
  \bibinfo{pages}{737} (\bibinfo{year}{1967}), ISSN \bibinfo{issn}{0031899X},
  \urlprefix\url{https://journals.aps.org/pr/abstract/10.1103/PhysRev.154.737}.

\bibitem[{\citenamefont{Konopka}(1967)}]{Konopka1967}
\bibinfo{author}{\bibfnamefont{J.}~\bibnamefont{Konopka}},
  \bibinfo{journal}{Physics Letters A} \textbf{\bibinfo{volume}{26}},
  \bibinfo{pages}{29} (\bibinfo{year}{1967}), ISSN \bibinfo{issn}{0375-9601}.

\bibitem[{\citenamefont{Mittag et~al.}(2019)\citenamefont{Mittag, Karalic, Lei,
  Thomas, Tuaz, Hatke, Gardner, Manfra, Ihn, and Ensslin}}]{Mittag2019}
\bibinfo{author}{\bibfnamefont{C.}~\bibnamefont{Mittag}},
  \bibinfo{author}{\bibfnamefont{M.}~\bibnamefont{Karalic}},
  \bibinfo{author}{\bibfnamefont{Z.}~\bibnamefont{Lei}},
  \bibinfo{author}{\bibfnamefont{C.}~\bibnamefont{Thomas}},
  \bibinfo{author}{\bibfnamefont{A.}~\bibnamefont{Tuaz}},
  \bibinfo{author}{\bibfnamefont{A.~T.} \bibnamefont{Hatke}},
  \bibinfo{author}{\bibfnamefont{G.~C.} \bibnamefont{Gardner}},
  \bibinfo{author}{\bibfnamefont{M.~J.} \bibnamefont{Manfra}},
  \bibinfo{author}{\bibfnamefont{T.}~\bibnamefont{Ihn}}, \bibnamefont{and}
  \bibinfo{author}{\bibfnamefont{K.}~\bibnamefont{Ensslin}},
  \bibinfo{journal}{Physical Review B} \textbf{\bibinfo{volume}{100}},
  \bibinfo{pages}{075422} (\bibinfo{year}{2019}), ISSN
  \bibinfo{issn}{24699969}, \eprint{1906.01995}.

\bibitem[{\citenamefont{Kjaergaard et~al.}(2017)\citenamefont{Kjaergaard,
  Suominen, Nowak, Akhmerov, Shabani, Palmstr{\o}m, Nichele, and
  Marcus}}]{Kjaergaard2017}
\bibinfo{author}{\bibfnamefont{M.}~\bibnamefont{Kjaergaard}},
  \bibinfo{author}{\bibfnamefont{H.~J.} \bibnamefont{Suominen}},
  \bibinfo{author}{\bibfnamefont{M.~P.} \bibnamefont{Nowak}},
  \bibinfo{author}{\bibfnamefont{A.~R.} \bibnamefont{Akhmerov}},
  \bibinfo{author}{\bibfnamefont{J.}~\bibnamefont{Shabani}},
  \bibinfo{author}{\bibfnamefont{C.~J.} \bibnamefont{Palmstr{\o}m}},
  \bibinfo{author}{\bibfnamefont{F.}~\bibnamefont{Nichele}}, \bibnamefont{and}
  \bibinfo{author}{\bibfnamefont{C.~M.} \bibnamefont{Marcus}},
  \bibinfo{journal}{Physical Review Applied} \textbf{\bibinfo{volume}{7}},
  \bibinfo{pages}{034029} (\bibinfo{year}{2017}), ISSN
  \bibinfo{issn}{23317019}, \eprint{1607.04164},
  \urlprefix\url{https://journals.aps.org/prapplied/abstract/10.1103/PhysRevApplied.7.034029}.

\bibitem[{\citenamefont{{Kittel C}}(1996)}]{Kittel1996}
\bibinfo{author}{\bibnamefont{{Kittel C}}}, \bibinfo{journal}{John Wiley \&
  Sons, New York}  (\bibinfo{year}{1996}).

\bibitem[{\citenamefont{Baumgartner et~al.}(2021)\citenamefont{Baumgartner,
  Fuchs, Fr{\'{e}}sz, Reinhardt, Gronin, Gardner, Manfra, Paradiso, and
  Strunk}}]{Baumgartner2021}
\bibinfo{author}{\bibfnamefont{C.}~\bibnamefont{Baumgartner}},
  \bibinfo{author}{\bibfnamefont{L.}~\bibnamefont{Fuchs}},
  \bibinfo{author}{\bibfnamefont{L.}~\bibnamefont{Fr{\'{e}}sz}},
  \bibinfo{author}{\bibfnamefont{S.}~\bibnamefont{Reinhardt}},
  \bibinfo{author}{\bibfnamefont{S.}~\bibnamefont{Gronin}},
  \bibinfo{author}{\bibfnamefont{G.~C.} \bibnamefont{Gardner}},
  \bibinfo{author}{\bibfnamefont{M.~J.} \bibnamefont{Manfra}},
  \bibinfo{author}{\bibfnamefont{N.}~\bibnamefont{Paradiso}}, \bibnamefont{and}
  \bibinfo{author}{\bibfnamefont{C.}~\bibnamefont{Strunk}},
  \bibinfo{journal}{Physical Review Letters} \textbf{\bibinfo{volume}{126}},
  \bibinfo{pages}{037001} (\bibinfo{year}{2021}), ISSN
  \bibinfo{issn}{10797114}, \eprint{2007.08371},
  \urlprefix\url{https://journals.aps.org/prl/abstract/10.1103/PhysRevLett.126.037001}.

\bibitem[{\citenamefont{{V. L. Bonch-Bruevitch and S. G.
  Kalashnikov}}(1990)}]{Bonch1990}
\bibinfo{author}{\bibnamefont{{V. L. Bonch-Bruevitch and S. G. Kalashnikov}}},
  \emph{\bibinfo{title}{{The Physics of Semiconductors}}}
  (\bibinfo{publisher}{Nauka Press}, \bibinfo{year}{1990}).

\bibitem[{\citenamefont{Suominen
  et~al.}(2017{\natexlab{b}})\citenamefont{Suominen, Danon, Kjaergaard,
  Flensberg, Shabani, Palmstr{\o}m, Nichele, and Marcus}}]{Suominen2017a}
\bibinfo{author}{\bibfnamefont{H.~J.} \bibnamefont{Suominen}},
  \bibinfo{author}{\bibfnamefont{J.}~\bibnamefont{Danon}},
  \bibinfo{author}{\bibfnamefont{M.}~\bibnamefont{Kjaergaard}},
  \bibinfo{author}{\bibfnamefont{K.}~\bibnamefont{Flensberg}},
  \bibinfo{author}{\bibfnamefont{J.}~\bibnamefont{Shabani}},
  \bibinfo{author}{\bibfnamefont{C.~J.} \bibnamefont{Palmstr{\o}m}},
  \bibinfo{author}{\bibfnamefont{F.}~\bibnamefont{Nichele}}, \bibnamefont{and}
  \bibinfo{author}{\bibfnamefont{C.~M.} \bibnamefont{Marcus}},
  \bibinfo{journal}{Physical Review B} \textbf{\bibinfo{volume}{95}},
  \bibinfo{pages}{035307} (\bibinfo{year}{2017}{\natexlab{b}}), ISSN
  \bibinfo{issn}{24699969}, \eprint{1611.00190},
  \urlprefix\url{https://journals.aps.org/prb/abstract/10.1103/PhysRevB.95.035307}.

\bibitem[{\citenamefont{Zeldov et~al.}(1994)\citenamefont{Zeldov, Clem,
  McElfresh, and Darwin}}]{Zeldov1994}
\bibinfo{author}{\bibfnamefont{E.}~\bibnamefont{Zeldov}},
  \bibinfo{author}{\bibfnamefont{J.~R.} \bibnamefont{Clem}},
  \bibinfo{author}{\bibfnamefont{M.}~\bibnamefont{McElfresh}},
  \bibnamefont{and} \bibinfo{author}{\bibfnamefont{M.}~\bibnamefont{Darwin}},
  \bibinfo{journal}{Physical Review B} \textbf{\bibinfo{volume}{49}},
  \bibinfo{pages}{9802} (\bibinfo{year}{1994}), ISSN \bibinfo{issn}{01631829},
  \urlprefix\url{https://journals.aps.org/prb/abstract/10.1103/PhysRevB.49.9802}.

\bibitem[{\citenamefont{Merservey and Schwartz}(1969)}]{Merservey1969}
\bibinfo{author}{\bibfnamefont{R.}~\bibnamefont{Merservey}} \bibnamefont{and}
  \bibinfo{author}{\bibfnamefont{B.~B.} \bibnamefont{Schwartz}}, in
  \emph{\bibinfo{booktitle}{Superconductivity}}, edited by
  \bibinfo{editor}{\bibfnamefont{R.~D.} \bibnamefont{Parks}}
  (\bibinfo{publisher}{Marcel Dekker, New York}, \bibinfo{year}{1969}), p.
  \bibinfo{pages}{126}.

\bibitem[{\citenamefont{Tinkham}(1996)}]{Tinkham1996}
\bibinfo{author}{\bibfnamefont{M.}~\bibnamefont{Tinkham}},
  \emph{\bibinfo{title}{{Introduction to superconductivity}}}
  (\bibinfo{publisher}{McGraw-Hill, Inc.}, \bibinfo{year}{1996}),
  \bibinfo{edition}{2nd} ed., ISBN \bibinfo{isbn}{0-07-064878-6}.

\bibitem[{\citenamefont{Gubin et~al.}(2005)\citenamefont{Gubin, Il'in,
  Vitusevich, Siegel, and Klein}}]{Gubin2005}
\bibinfo{author}{\bibfnamefont{A.~I.} \bibnamefont{Gubin}},
  \bibinfo{author}{\bibfnamefont{K.~S.} \bibnamefont{Il'in}},
  \bibinfo{author}{\bibfnamefont{S.~A.} \bibnamefont{Vitusevich}},
  \bibinfo{author}{\bibfnamefont{M.}~\bibnamefont{Siegel}}, \bibnamefont{and}
  \bibinfo{author}{\bibfnamefont{N.}~\bibnamefont{Klein}},
  \bibinfo{journal}{Physical Review B} \textbf{\bibinfo{volume}{72}},
  \bibinfo{pages}{064503} (\bibinfo{year}{2005}), ISSN
  \bibinfo{issn}{1550235X},
  \urlprefix\url{https://journals.aps.org/prb/abstract/10.1103/PhysRevB.72.064503}.

\bibitem[{\citenamefont{Pearl}(1964)}]{Pearl1964}
\bibinfo{author}{\bibfnamefont{J.}~\bibnamefont{Pearl}},
  \bibinfo{journal}{Applied Physics Letters} \textbf{\bibinfo{volume}{5}},
  \bibinfo{pages}{65} (\bibinfo{year}{1964}), ISSN \bibinfo{issn}{0003-6951},
  \urlprefix\url{https://aip.scitation.org/doi/abs/10.1063/1.1754056}.

\bibitem[{\citenamefont{Rodan-Legrain et~al.}(2021)\citenamefont{Rodan-Legrain,
  Cao, Park, de~la Barrera, Randeria, Watanabe, Taniguchi, and
  Jarillo-Herrero}}]{Rodan2021}
\bibinfo{author}{\bibfnamefont{D.}~\bibnamefont{Rodan-Legrain}},
  \bibinfo{author}{\bibfnamefont{Y.}~\bibnamefont{Cao}},
  \bibinfo{author}{\bibfnamefont{J.~M.} \bibnamefont{Park}},
  \bibinfo{author}{\bibfnamefont{S.~C.} \bibnamefont{de~la Barrera}},
  \bibinfo{author}{\bibfnamefont{M.~T.} \bibnamefont{Randeria}},
  \bibinfo{author}{\bibfnamefont{K.}~\bibnamefont{Watanabe}},
  \bibinfo{author}{\bibfnamefont{T.}~\bibnamefont{Taniguchi}},
  \bibnamefont{and}
  \bibinfo{author}{\bibfnamefont{P.}~\bibnamefont{Jarillo-Herrero}},
  \bibinfo{journal}{Nature Nanotechnology} \textbf{\bibinfo{volume}{16}},
  \bibinfo{pages}{769} (\bibinfo{year}{2021}), ISSN \bibinfo{issn}{1748-3395},
  \eprint{2011.02500},
  \urlprefix\url{https://www.nature.com/articles/s41565-021-00894-4}.

\end{thebibliography}

\end{document}